\documentclass[5p,times]{elsarticle}
 
\usepackage{graphicx} 
\usepackage{acronym}
\usepackage{amsmath}
\usepackage{amssymb}
\usepackage{array}
\usepackage{balance}
\usepackage{pifont}
\usepackage{subcaption}
\usepackage{wasysym} 
\usepackage[hang,flushmargin]{footmisc}
\usepackage{multirow} 
\usepackage{wasysym} 
\usepackage{makecell} 
\usepackage{booktabs} 
\usepackage{tikz} 
\usepackage{hyperref}
\newcommand*\emptycirc[1][1ex]{\tikz\draw[thick] (0,0) circle (#1);} 

\newcommand*\fullcirc[1][1ex]{\tikz\fill (0,0) circle (#1);} 

\usepackage{makecell} 

\acrodef{UAV}{Unmanned Aerial Vehicle}
\acrodef{SDR}{Software-Defined Radio}
\acrodef{RF}{Radio Frequency}
\acrodef{RFF}{Radio Frequency Fingerprinting}
\acrodef{LEO}{Low-Earth Orbit}
\acrodef{BPSK}{Binary Phase Shift Keying}
\acrodef{QPSK}{Quadrature Phase Shift Keying}
\acrodef{DQPSK}{Differential \ac{QPSK}}
\acrodef{IQ}{In-Phase Quadrature}
\acrodef{SNR}{Signal-to-Noise Ratio}
\acrodef{COTS}{Commercial-Off-The-Shelf}
\acrodef{MSE}{Mean Squared Error}
\acrodef{CSI}{Channel State Information}
\acrodef{RFF}{Radio Frequency Fingerprinting}
\acrodef{GPS}{Global Positioning System}
\acrodef{GNSS}{Global Navigation Satellite System}
\acrodef{PHY}{Physical}
\acrodef{AE}{Autoencoder}
\acrodef{LNA}{Low-Noise Amplifier}
\acrodef{AUC}{Area Under-the-Curve}
\acrodef{FCDD}{Fully Convolutional Data Descriptor}
\acrodef{RBM}{Restricted Boltzmann Machine}
\acrodef{HPC}{High-Performance Computing}
\acrodef{ROC}{Receiver Operating Characteristic}
\acrodef{IoT}{Internet of Things}
\acrodef{LSTM}{Long Short-Term Memory}

\newcolumntype{P}[1]{>{\centering\arraybackslash}p{#1}}

\def\BibTeX{{\rm B\kern-.05em{\sc i\kern-.025em b}\kern-.08em
    T\kern-.1667em\lower.7ex\hbox{E}\kern-.125emX}}

\journal{Computer Networks}
\begin{document}
\begin{frontmatter}

\title{Detection of Aerial Spoofing Attacks to LEO Satellite Systems via Deep Learning}


\tnotetext[t1]{This work has been partially supported by the INTERSECT project, Grant No. NWA.1162.18.301, funded by the Netherlands Organization for Scientific Research (NWO). Any opinions, findings, conclusions, or recommendations expressed in this work are those of the author(s) and do not necessarily reflect the views of NWO. Moreover, this publication was made possible by the NPRP12C-0814-190012-SP165 awards from the Qatar National Research Fund (a member of Qatar Foundation).}

\author[1]{Jos Wigchert}
\ead{j.wigchert@student.tue.nl}
\author[1]{Savio Sciancalepore\corref{cor1}}
\ead{s.sciancalepore@tue.nl}
\author[2]{Gabriele Oligeri}
\ead{goligeri@hbku.edu.qa}
\cortext[cor1]{Corresponding author}

\affiliation[1]{organization={Eindhoven University of Technology (TU/e)},
city={Eindhoven},
country={Netherlands}}

\affiliation[2]{organization={Hamad Bin Khalifa University (HBKU), College of Science and Engineering (CSE)},
city={Doha},
country={Qatar}}

\begin{abstract}
    Detecting spoofing attacks to Low-Earth-Orbit (LEO) satellite systems is a cornerstone to assessing the authenticity of the received information and guaranteeing robust service delivery in several application domains. The solutions available today for spoofing detection either rely on additional communication systems, receivers, and antennas, or require mobile deployments. Detection systems working at the Physical (PHY) layer of the satellite communication link also require time-consuming and energy-hungry training processes on all satellites of the constellation, and rely on the availability of spoofed data, which are often challenging to collect. Moreover, none of such contributions investigate the feasibility of aerial spoofing attacks launched via drones operating at various altitudes.
    In this paper, we first show experimentally the viability and effectiveness of spoofing attacks to LEO satellite systems using aerial attackers deployed on drones. We also propose a new spoofing detection technique, relying on pre-processing raw physical-layer signals into images and then applying anomaly detection on such images via autoencoders. We validate our solution through an extensive measurement campaign involving the deployment of an actual spoofer (Software-Defined Radio) installed on a drone and injecting rogue IRIDIUM messages while flying at different altitudes with various movement patterns. Our results demonstrate that the proposed technique can reliably detect LEO spoofing attacks launched at different altitudes, while state-of-the-art competing approaches simply fail. We also release the collected data as open source, fostering further research on satellite security.
\end{abstract}


\end{frontmatter}

\section{Introduction}
\label{sec:intro}

\textcolor{black}{\ac{LEO} satellite communication systems have recently gained renewed momentum in \ac{IoT} research and Industry, thanks to the enhanced efficiency, simplicity, and low cost in technology access~\cite{prol2022_access}~\cite{zhao2025_tmc}. Several commercial \ac{IoT}-based applications are also benefiting from the integration of such a technology, with new powerful satellite-based services emerging like real-time data collection, predictive maintenance, asset tracking, cold-chain management, and resource exploration, to name a few, and new communication paradigms appearing, e.g., Mobile Edge Computing~\cite{zhao2024_iotj}.}

At the same time, the existence of a satellite communication link also increases the threat surface of such users. Among the various cybersecurity threats affecting LEO satellite systems, \emph{spoofing attacks}, a.k.a. \emph{wireless impersonation}, exploit the absence (or compromise) of message authentication techniques to inject rogue messages on the satellite link~\cite{merwe2018_enc}. Such messages are crafted to be indistinguishable from legitimate ones, being received and processed by commercial satellite receivers, leading to successful attacks~\cite{salkield2023_wisec},~\cite{cyr2023_spacesec}.  

Many solutions are available in the literature for the detection of satellite spoofing~\cite{yue2023_comst},~\cite{tedeschi2022_comnet}. However, they require several communication systems, receivers and antennas, or mobile systems distributed in a large area, being quite expensive to deploy for many application domains. When considering standalone non-mobile systems, the literature suggests the use of transmitter identification approaches, as in~\cite{oligeri2023tifs}, or the usage of fading phenomena to detect spoofing, as in~\cite{sadighian2024ccnc} and~\cite{oligeri2024sac}. However, these solutions require binary classification, building on top of time- and resource-consuming training processes involving both legitimate and spoofed signals, which are often either impractical or unlawful to acquire. Some scientific contributions such as~\cite{yue2023_comst} and~\cite{salkield2023_wisec} identified the feasibility of spoofing attacks to LEO satellite systems, while some others identified the threats posed by aerial spoofing platforms, such as Kumar et al.~\cite{kumar2023_authorea}. However, none of the contributions available in the literature for satellite spoofing detection provide solutions that do not require the availability of attack data during system training. Also, none of these works perform real-world experimentation with aerial spoofing attacks carried out by drones hovering or flying at various altitudes. Using such a strategy, the adversary could mimic better the satellite downlink to avoid detection and maximize damage.

{\bf Contribution.} In this paper, we provide the following contributions: (i) we demonstrate the technical feasibility of \emph{aerial spoofing} attacks to LEO satellite systems, where an adversary uses drones operating at various altitudes (beyond Line-of-Sight) to stealthily inject spoofed messages to a target satellite receiver; (ii) we demonstrate that existing solutions to detect LEO spoofing attacks fail to discover such attacks, misclassifying spoofed messages as legitimate ones; (iii) we design a new solution for aerial spoofing attacks to LEO satellite systems, pre-processing raw IQ samples into images and using a sparse \ac{AE} to reliably detect anomalies in the received signals; and finally, (iv) through extensive real-world experimentation involving actual drones and \acp{SDR}, we demonstrate the robustness and effectiveness of our proposed solution and its superior performances against the state of the art. To the best of our knowledge, this paper is the first to consider real-world aerial spoofing attacks to LEO satellite systems and to propose an ad-hoc pre-processing technique of physical-layer information enabling attack detection via sparse \ac{AE}. Although similar techniques have been applied for identifying specific satellites, e.g., in~\cite{oligeri2023tifs}, we demonstrate experimentally that satellite link fingerprinting is better suited for detecting spoofing attacks compared to identifying single satellites.
We also release the dataset collected for this work at~\cite{dataset}, enabling further research on LEO satellite links modeling, monitoring, and security.

We highlight that this paper extends our previous research in~\cite{sadighian2024ccnc} and ~\cite{oligeri2024sac} through the following new contributions:
\begin{itemize}
    \item We demonstrate the technical feasibility of aerial spoofing attacks to LEO satellite systems, using a drone operating at various altitudes to inject replayed IRIDIUM messages to a receiver on the ground.
    \item While previous research leveraged binary classification for spoofing detection, in this paper, we demonstrate the feasibility of detecting spoofing attacks to LEO satellite systems using a pre-processing technique transforming physical-layer information into images, enabling spoofing detection through one-class classification via autoencoders, thus requiring the availability of only regular communication samples at training time.
    \item While previous research demonstrated the effectiveness of fading fingerprinting by only considering terrestrial attacks, in this paper, we consider a novel attack strategy, i.e., aerial spoofing attacks launched through drones.
    \item We collect a brand new real-world dataset, including benign satellite data and actual spoofing attacks carried out using \acp{SDR} and drones, and release this dataset as open source to the research community at~\cite{dataset}.
    \item We compare the performance of our newly proposed approach against preliminary solutions proposed in~\cite{oligeri2024sac} and \cite{oligeri2023tifs}, demonstrating the superiority of our approach in terms of attack detection accuracy while also requiring fewer samples than competing solutions. 
\end{itemize}

{\bf Roadmap.} This paper is organized as follows. Sec.~\ref{sec:background} introduces the preliminaries, Sec.~\ref{sec:related} reviews related work, Sec.~\ref{sec:sys_advModel} introduces the system and adversary model, Sec.~\ref{sec:methodology} provides the details of our solution, Sec.~\ref{sec:data_collection} describes our data collection, Sec.~\ref{sec:performance} provides the results, Sec.~\ref{sec:discussion} discusses implications and limitations of our research, and finally, Sec.~\ref{sec:conclusion} draws the conclusion.

\section{Background}
\label{sec:background}

This section introduces preliminary concepts, i.e., digital modulation principles, used as an input to detecting spoofing attacks, the IRIDIUM satellite infrastructure, used for our experimental testing phase, and \acp{AE}, used as part of our detection solution.\\\\
\noindent
{\bf Digital (De-)Modulation.} Modern wireless communication technologies use digital modulation to make bit-strings suitable for transmission over the \ac{RF} link. Such signals are typically expressed in the form of a complex number $I+jQ$, namely \ac{IQ} samples, being $I$ the \emph{In-Phase} and $Q$ the \emph{Quadrature} component. These components are (90 degrees) phase-shifted by the transmitter before transmission so that they do not interfere with each other while propagating over the wireless channel. The original \ac{IQ} value (and bitstream) can be recovered by the receiver by multiplying the received signal via two orthogonal tones~\cite{rappaport2024_book}. The number of possible symbols $s$ defined by the modulation scheme, i.e., $log_{2}{s}$ bits/sec, impacts the throughput of the communication link. In addition, $s$ defines the tolerance of the communication link to noise. In fact, the \ac{IQ} symbols at the receiver are affected by noise $n$, introduced jointly by the electronics of the transmitter and the communication channel. The receiver simply maps the received \ac{IQ} symbol $x = I + jQ + n$ to the closer expected one, i.e., the symbol characterized by the minimum error. The higher the noise power, the higher the probability that a symbol is distorted enough to get closer to another one, causing a decoding error. Thus, communication links affected by high noise, such as satellite links, use lower-order schemes, e.g., \ac{BPSK} ($s=2$) and \ac{QPSK} ($s=4$), trading off throughput with robustness to noise. Also, note that \acp{COTS} cannot easily control the value of the \ac{IQ} samples. Indeed, the \ac{IQ} samples received by a device are affected by both manufacturing inaccuracies in the transmitter hardware, which are inherent to such devices, and fading phenomena caused by the wireless channel~\cite{shawabka2020_infocom}. In our paper, we analyze the distortion affecting the \ac{IQ} samples received by a target user to detect spoofing attacks (see Sec.~\ref{sec:methodology}).\\

\noindent
{\bf IRIDIUM.} 
IRIDIUM is a \ac{LEO} satellite constellation operated by IRIDIUM Communications Inc, offering global satellite phone and data coverage currently to more than 2 millions users \cite{iridium}, \cite{iridium_users}. IRIDIUM enables users to communicate from anywhere on Earth, including remote areas where traditional terrestrial communication networks typically cannot provide connectivity.
The constellation includes $66$ \ac{LEO} satellites orbiting at $800$~km all over the Earth's surface and moving from north to south~\cite{oligeri2020_wisec}. 
The satellites transmit messages in the frequency band $[1,618.75 - 1,626.5]$~MHz using the modulation scheme \ac{DQPSK}, which can be received through dedicated satellite receiving stations provided by authorized resellers such as Motorola and Kyocera.
IRIDIUM features two main types of communication channels, i.e., \emph{system overhead} and \emph{bearer service} channels~\cite{pratt1999_comst}. Some channels are authenticated, while others are not authenticated and not encrypted, allowing for spoofing attacks. Unauthenticated and unencrypted channels, e.g., IRIDIUM Ring Alert (IRA), are used to broadcast information about a specific satellite. The information delivered by satellites on such channels can be detected and decoded using a \ac{SDR}, a generic active antenna, and a software development kit like \mbox{\emph{gr-iridium}}~\cite{gr_iridium}.  
IRIDIUM satellites serve primarily as a communication system. However, the network can also be used for positioning purposes, e.g., through systems like \emph{DDK Positioning}~\cite{dmackay_2021}. IRIDIUM also features a dedicated IRIDIUM IoT service, specifically dedicated to Machine-to-Machine (M2M) applications like transportation, agriculture, oil and gas, utilities, and construction, among many others~\cite{iridium_iot}.
In this paper, we use IRIDIUM as the reference \ac{LEO} satellite constellation to test spoofing attacks because of its popularity, number of users, and availability of an open-source reception module for \acp{SDR} to receive real messages, enabling our experimental data collection and analysis. \\

\noindent
{\bf Autoencoders (AE).} 
\acp{AE} are special types of neural network architectures used for unsupervised learning tasks, including anomaly detection and dimensionality reduction~\cite{li2023_asc}. They feature an encoder, compressing the input data \( M 
\) into a lower-dimensional representation, and a decoder, that reconstructs the original data from such compressed representation.
%
The encoder function \( G_\theta \) maps the input image \( M \) to a latent representation \( z \in \mathbb{R}^d \), through a transformation of the type $z = G_\theta(M)$, where $\theta$ represents the encoder function with related parameters and \( d \) is the dimensionality of the latent space, typically much smaller than the original image dimensions \( M_W \times M_H \)~\cite{sciancalepore2024_iotj}.
The decoder function \( F_\phi \) maps the latent representation back to the reconstructed image \( \hat{M} \), through the transformation $\hat{M} = F_\phi(z)$, being  $( \phi)$ the decoder transfer function with specific decoder parameters.
\acp{AE} are trained to minimize the reconstruction error $\mathcal{L}$ between the original input image \( M \) and the reconstructed image \( \hat{M} \), measured using the \ac{MSE} loss function, as in Eq. \ref{eq:mse_loss}~\cite{zhou2022_icml}.
\begin{equation}
    \label{eq:mse_loss}    
    \mathcal{L}(\theta, \phi) = \frac{1}{M_W M_H} \sum_{i=1}^{M_W} \sum_{j=1}^{M_H} (M_{i,j} - \hat{M}_{i,j})^2. 
\end{equation}
Note that \acp{AE} can also be used as a one-class classification tool, i.e., they can be trained only on samples from one class and then used to detect anomalies as outliers. This is particularly useful when dealing with aerial spoofing attacks. Indeed, collecting attack samples (anomalies) is particularly challenging because of the equipment, technical difficulties required to assemble the experimental setup and launch the attack, and finally, law and safety regulations, which often prevent the collection of such data. Recent related work, such as the one by the authors in~\cite{vskvara2021_tnnls}, also identified experimentally the advantage of autoencoder-based models when a limited number of anomalous samples is available. We use this feature as part of our solution (see Sec.~\ref{sec:methodology}).

\begin{table*}
\small
    \centering
    \begin{tabular}{|c|c c c c c c c c|} 
        \hline
        {\bf Paper} & \makecell{ {\bf Location} \\ {\bf Independence}} & \makecell{ {\bf Single} \\ {\bf Receiver}} & \makecell{ {\bf Support for} \\ {\bf Static Receivers}} & {\bf No crypto} & \makecell { {\bf No Multiple} \\ {\bf Sources}} & \makecell{ \bf{ Applied on} \\ {\bf Satellites}} & \makecell{ \bf{No Binary} \\ {\bf Classification}}  & \makecell{ {\bf Aerial Adversary}} \\ [0.5ex] 
        \hline
        \cite{oligeri2024sac}   & $\fullcirc$  & $\fullcirc$  & $\fullcirc$  & $\fullcirc$ & $\fullcirc$ & $\fullcirc$ & $\emptycirc$ & $\emptycirc$ \\ 
        \cite{kumar2023_authorea} & $\fullcirc$ & $\fullcirc$ & $\fullcirc$ & $\fullcirc$ & $\fullcirc$ & $\fullcirc$ & $\emptycirc$ & $\emptycirc$ \\ 
        \cite{wang2015_wcnc}                   & $\emptycirc$ & $\fullcirc$  & $\fullcirc$  & $\fullcirc$ & $\fullcirc$ & $\emptycirc$ & $\emptycirc$ & $\emptycirc$ \\ 
        \cite{oligeri2020_wisec}                  & $\fullcirc$  & $\fullcirc$  & $\fullcirc$  & $\fullcirc$ & $\emptycirc$ & $\fullcirc$ & $\emptycirc$ & $\emptycirc$ \\
        \cite{fourandeh2020_wisec} & $\fullcirc$  & $\fullcirc$  & $\fullcirc$  & $\fullcirc$ & $\fullcirc$ & $\fullcirc$ & $\emptycirc$ & $\emptycirc$ \\
        \cite{sun2021_tim} & $\emptycirc$  & $\fullcirc$  & $\fullcirc$  & $\fullcirc$ & $\fullcirc$ & $\fullcirc$ & $\fullcirc$ & $\emptycirc$ \\
        \cite{chapre2015_percom}                 & $\emptycirc$ & $\emptycirc$ & $\fullcirc$  & $\emptycirc$ & $\fullcirc$ & $\emptycirc$ & $\emptycirc$ & $\emptycirc$ \\ 
        \cite{psiaki2013_taes}              & $\fullcirc$  & $\emptycirc$ & $\fullcirc$  & $\fullcirc$ & $\emptycirc$ & $\fullcirc$ & $\fullcirc$ & $\emptycirc$ \\ 
        \cite{hua2018_infocom}                  & $\fullcirc$  & $\fullcirc$  & $\fullcirc$  & $\fullcirc$ & $\fullcirc$ & $\emptycirc$ & $\emptycirc$ & $\emptycirc$ \\ 
        \cite{broumandan2012_plans}            & $\fullcirc$  & $\fullcirc$  & $\emptycirc$ & $\emptycirc$ & $\fullcirc$ & $\fullcirc$ & $\fullcirc$ & $\emptycirc$ \\ 
        \cite{abdrabou2022_ojcs} & $\emptycirc$ & $\fullcirc$ & $\fullcirc$ & $\fullcirc$ & $\fullcirc$ & $\fullcirc$ & $\fullcirc$ & $\emptycirc$ \\
        \cite{abdrabou2024_tcc} & $\emptycirc$ & $\fullcirc$ & $\fullcirc$ & $\fullcirc$ & $\fullcirc$ & $\fullcirc$ & $\fullcirc$ & $\emptycirc$ \\
        \hline
        Ours                    & $\fullcirc$  & $\fullcirc$  & $\fullcirc$  & $\fullcirc$ & $\fullcirc$ & $\fullcirc$ & $\fullcirc$ & $\fullcirc$ \\ 
        \hline
    \end{tabular}
    \caption{Related work on Spoofing Detection.
    }
    \label{tab:paper_comparison}
\end{table*}

\section{Related Work}
\label{sec:related}

Spoofing of satellite messages is usually associated with \ac{GNSS} spoofing, i.e., the falsification of positioning and navigation messages emitted by systems such as GPS, GALILEO, and GLONASS~\cite{schmidt2016_csur}. Although secure versions of such communication technologies exist and are used by the military, for backwards-compatibility reasons, civilian systems mostly use their former versions, which are insecure by design. In this context, He et al. in~\cite{he2016_wcom} also identified that, similarly to GNSS technologies like GPS, LEO satellite networks typically use a dedicated downlink pilot channel for broadcasting channel status, user management information, call information, etc., as exemplified by Iridium. Attackers can imitate the dedicated pilot channel to broadcast false information to legitimate users, causing network paralysis. However, they did not provide valid countermeasures for such attacks.
One common strategy to detect spoofing attacks involves the exploration of alternative sources, such as military \ac{GNSS} signals \cite{psiaki2013_taes}, terrestrial communication systems~\cite{oligeri2022_comnet}, or the integration of data from additional \ac{GNSS} constellations, such as Iridium \cite{oligeri2020_wisec}. 
These methodologies utilize the correlation between the information collected in distinct channels to assess the authenticity of \ac{GNSS} signals. However, they depend on the availability and authenticity of an alternative source of information. Some other proposals, such as~\cite{chapre2015_percom}, use multiple receivers to detect spoofing attacks, leveraging spatial and channel diversity and the use of \ac{CSI}-based fingerprints of the received signal. This approach has also been used by the authors in~\cite{wang2015_wcnc} and~\cite{hua2018_infocom} for WiFi signals, and is overall very promising for indoor spoofing detection. However, as shown in Sec.~\ref{sec:results}, it hardly applies to GNSS spoofing due to high channel noise. 
To create channel diversity without several receivers, the authors in~\cite{broumandan2012_plans} use a mobile \ac{GPS} receiver. 
However, such solutions require mobile receivers and usually incur a non-negligible delay, necessary to actually move from one location to another.
Recent popular methods to detect (GNSS) spoofing attacks involve the use of \ac{RFF}, i.e., the creation of a transmitter fingerprint that helps distinguish legitimate from rogue ones. RFF has been applied successfully for terrestrial communication technologies~\cite{alhazbi2023_acsac},
LEO~\cite{oligeri2023tifs},~\cite{smailes2023_ccs},~\cite{solenthaler2025_spacesec}, and GPS satellites~\cite{fourandeh2020_wisec}. However, it requires significant training time and does not scale up with the number of legitimate satellites, which is typically large for LEO constellations. Other proposals in the GNSS domain, such as the one by Borhani et al. in~\cite{borhani2024_jasp}, use GNSS-specific metrics, such as Cross Ambiguity Function (CAF), to detect spoofing attacks. Being such metrics specific to GNSS technologies, they cannot be used for LEO satellite spoofing detection.
Some more recent proposals, such as~\cite{sadighian2024ccnc},~\cite{oligeri2024sac}, and~\cite{sun2021_tim}, use \ac{IQ} samples to detect spoofing attacks affecting both the LEO and GNSS constellations. While the authors in~\cite{sun2021_tim} detect anomalies in the Quadrature component of IQ signals, the proposals in~\cite{sadighian2024ccnc} and~\cite{oligeri2024sac} apply binary classification to both I and Q components to detect spoofing and require knowledge of both spoofed and legitimate samples. Binary classification is also required by Kumar et al. in~\cite{kumar2023_authorea}, which applied \ac{LSTM} to noise data via simulations to detect aerial spoofing. It is also worth mentioning the contributions by Abdrabou et al. in~\cite{abdrabou2022_ojcs} and~~\cite{abdrabou2024_tcc}, using Doppler frequency shift (DS) and received power (RP) features to detect spoofing attacks to satellite systems. However, none of these solutions take into account spoofing attacks carried out by aerial attackers, such as drones.\\
Thus, as summarized in Table~\ref{tab:paper_comparison}, the current literature misses actual real-world tests involving spoofing attacks from aerial adversaries. Moreover, existing methods cannot provide a scalable mechanism independent of the number of satellites that comprise the LEO constellation. In Sec.~\ref{sec:results}, we also compare our solution to the solutions proposed in \cite{oligeri2024sac} and \cite{oligeri2023tifs}, demonstrating superior performance. 

\section{System and Adversary Model}
\label{sec:sys_advModel}

\begin{figure}[h!]
    \centering
    \includegraphics[width=\columnwidth]{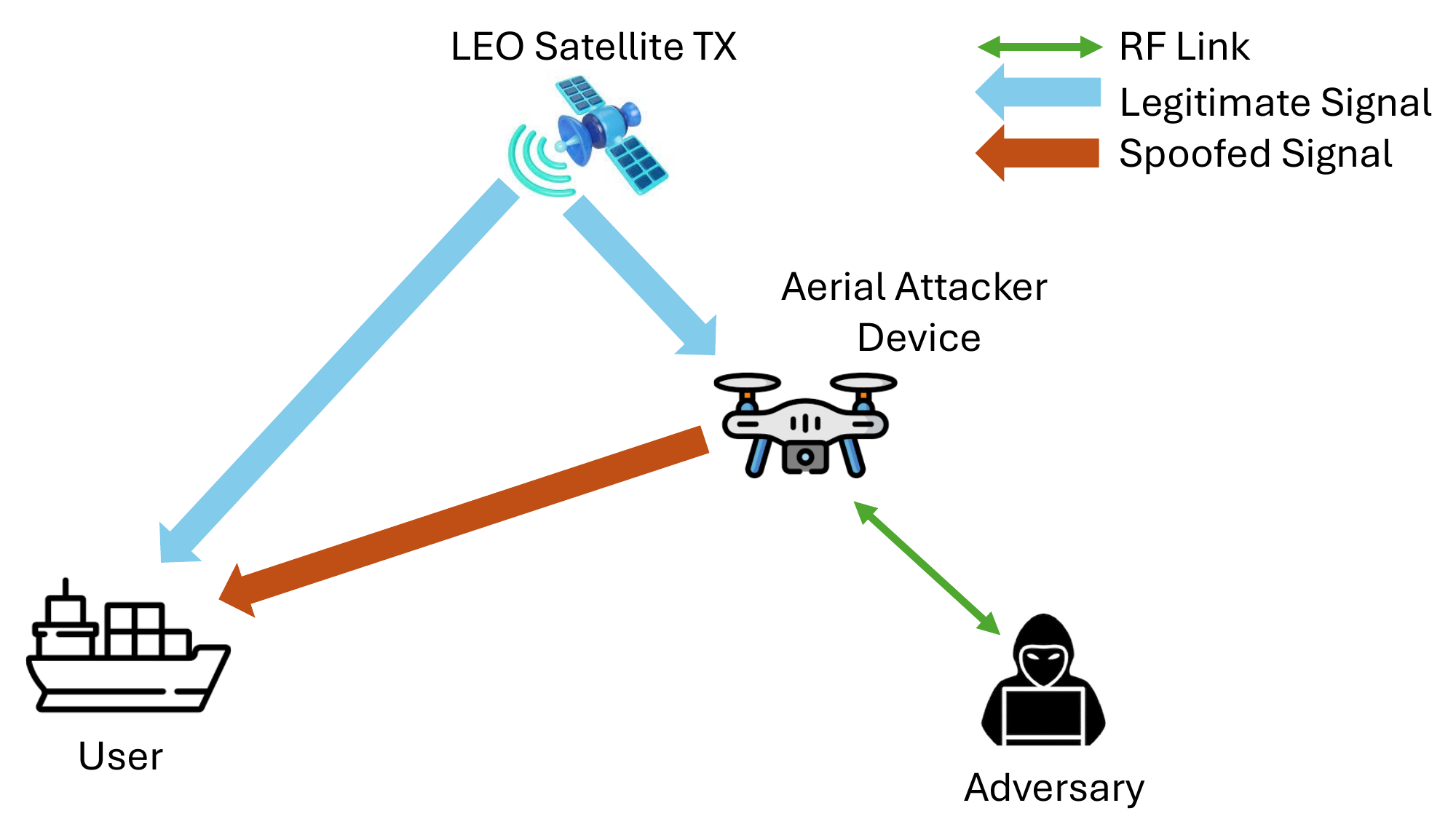}
    \caption{System and adversary model. An adversary deploys a drone to spoof LEO satellite messages delivered by legitimate satellites at a target receiver.}
    \label{fig:scenario}
\end{figure}

Figure~\ref{fig:scenario} illustrates the system and adversary models considered in this work.

Our reference scenario includes three relevant entities: (i) the satellite transmitter, (ii) the user equipped with a satellite receiver, and (iii) the adversary. 

Without loss of generality, we consider a satellite transmitter located in the \ac{LEO}, emitting \ac{RF} messages according to the IRIDIUM communication system, as discussed in Sec.~\ref{sec:background}. Thus, legitimate signals emitted by the satellite follow the modulation scheme \ac{QPSK} with differential encoding, a.k.a. \ac{DQPSK}, and are transmitted in the frequency band $[1,618.725 - 1,626.5]$~MHz.
The user features a receiving device able to receive and decode IRIDIUM satellite messages, e.g., for communication in hard-to-reach locations. We consider a receiver deployed outdoors for maximum satellite visibility. We do not make any additional assumptions on the deployment of the receiver, which can be either static or mobile. However, we consider that the user does not feature any additional Internet connection than the one possibly provided by the LEO satellite system. Thus, the user cannot resort to any other service than the IRIDIUM communication system to verify the legitimacy of the information provided through such a system. This consideration applies to many IRIDIUM deployments, as the system is typically used to enable communications in remote and hard-to-reach locations. Moreover, note that we require raw physical-layer \ac{IQ} data to detect spoofing attacks. Such data are usually not available when using \acp{COTS}, while they are available by using \acp{SDR}, as we did in our experimental campaign. Thus, users willing to deploy our solution require a \ac{SDR} to collect the data aside to the IRIDIUM receiver provided by the service provider.

{\bf Adversary Model.} The adversary considered in this work, namely $\mathcal{A}$, aims to spoof satellite messages, making them appear legitimate to the user for processing. The adversary's objective is to induce the computation of a different position, velocity, and/or time (PVT) in a targeted civilian IRIDIUM receiver. This objective is feasible since many (LEO) satellite systems, including IRIDIUM, use unencrypted and unauthenticated messages. In addition, messages transmitted from LEO satellites typically arrive at the receiver with low \ac{SNR}, and they are easily overcome in terms of power by the ones injected by the adversary.  In practice, $\mathcal{A}$ can use a \ac{SDR} to both eavesdrop on legitimate messages delivered by legitimate IRIDIUM satellites, and inject new spoofed messages on the communication channel. In this paper, we specifically consider the scenario in which $\mathcal{A}$ uses an \ac{UAV}, a.k.a. drone, to launch replay (meaconing) and spoofing attacks. The use of a drone as a transmitting source makes the received signal (and the associated fading phenomenon) more similar to a satellite downlink, thus having higher chances to bypass (state-of-the-art) spoofing detection solutions based on fading fingerprinting, such as the ones in~\cite{sadighian2024ccnc} and~\cite{oligeri2024sac}.
The objective of this paper is to discriminate legitimate (benign) from spoofed (malicious) IRIDIUM messages through the analysis of the information available at the physical layer of the communication link. 

\section{Satellite Spoofing Detection}
\label{sec:methodology}

In this section, we describe the methodology used to detect spoofing attacks to LEO satellite communication systems. 
Our proposed methodology for spoofing detection is divided into two phases, i.e., \emph{image generation} from physical layer information, i.e., \ac{IQ} samples, (see Sec.~\ref{sec:image_generation}) and \emph{spoofing detection} (Sec.~\ref{sec:autoencoder_detection}). Tab.~\ref{tab:notation} summarizes our main notation.
\begin{table}[h]
\small
\centering
\begin{tabular}{|c|c|}
\hline
\textbf{Symbol} & \textbf{Description} \\
\hline
$I_k$ & In-phase component of the $k$-th IQ sample. \\
$Q_k$ & Quadrature component of the $k$-th IQ sample. \\
$N$ & Total number of IQ samples. \\
$P$ & Number of histogram bins along the $I$ axis. \\
$Q$ & Number of histogram bins along the $Q$ axis. \\
$b_{I,j}$ & $j$-th bin edge along the $I$ axis. \\
$b_{Q,j}$ & $j$-th bin edge along the $Q$ axis. \\
$H_{i,j}$ & Value of the histogram at bin $(i,j)$. \\
$M_{i,j}$ & Value of the image at pixel $(i,j)$. \\
$\mathbb{I}(\circ)$ & Indicator function. \\
$L$ & Number of neurons of the encoder of the \ac{AE}. \\
$S$ & Number of neurons of the decoder of the \ac{AE}. \\
$\mathbf{MSE_{train}}$ & Vector of \ac{MSE} values generated during \ac{AE} training. \\
$\tau$ & Threshold of the \ac{AE}. \\
$E(\circ)$ & Mean operator. \\
$\sigma(\circ)$ & Standard Deviation operator. \\
\hline
\end{tabular}
\caption{Notation and description.}
\label{tab:notation}
\end{table}

\subsection{Image Generation}
\label{sec:image_generation}

\begin{figure}
    \centering
    \includegraphics[width=\columnwidth]{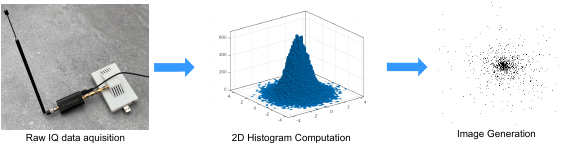}
    \caption{Graphical overview of the image generation \mbox{methodology} used in our solution. 
    }
    \label{fig:image_generation}
\end{figure}

Figure \ref{fig:image_generation} shows the image generation methodology used in our approach.
Our solution analyzes raw \ac{IQ} samples, collected through a device capable of extracting \ac{PHY}-layer data from the received signal, e.g., a \ac{SDR} or a spectrum analyzer. We divide the stream of \ac{IQ} samples into subsets (non-overlapping chunks) of $N$ samples, i.e., $\{(I_k, Q_k)\}_{k=1}^{N}$ (see Sec.~\ref{sec:results} for an evaluation of the impact of $N$ on the performance of our solution), and visualize them through an \ac{IQ} constellation diagram, with the I component on the x-axis and the Q component on the y-axis. Then we divide the diagram into a $P\times Q$ grid, with $P$ and $Q$ chosen according to the methodology used for spoofing detection (see Sec.~\ref{sec:autoencoder_detection}). We count how many samples fall into each tile of the grid, so generating a histogram $H$. Denoting with $\mathbf{b_{I}}$ and $\mathbf{b_Q}$ the set of edges of the tiles of the histogram along X and Y, respectively, the value of the histogram tile $H_{i,j}$ follows Eq. \ref{eq:hist_set}.
\begin{equation}
\label{eq:hist_set}
    H_{i,j} = \sum_{k=1}^{N} \mathbb{I}(b_{I,i} \leq I_k < b_{I,i+1}) \cdot \mathbb{I}(b_{Q,j} \leq Q_k < b_{Q,j+1}),
\end{equation}
where $\mathbb{I}$ is the indicator function, defined as per Eq. \ref{eq:indicator_function}.
\begin{equation}
\label{eq:indicator_function}
    \mathbb{I}(\circ) = 
    \begin{cases} 
    1 & \text{if } \circ \text{ evaluates to true,} \\
    0 & \text{otherwise.}
    \end{cases}
\end{equation}

We interpret each tile of the histogram as a pixel of a grey-scale image, where $0$ is black and $255$ is white. If $N>255$, it is possible that $H_{i,j}>255$; in such a case, we truncate such a value to $255$. Finally, we provide the resulting matrix (image) as input for the following spoofing detection phase. Note that, by converting IQ samples into an image, we provide the \ac{AE} with a structured representation of the RF signal, facilitating effective learning and anomaly detection.

\subsection{Spoofing Detection via Autoencoders}
\label{sec:autoencoder_detection}

We identify a spoofing attack using sparse \acp{AE}, by detecting anomalies in the images generated from the received RF signal due to different fading phenomena affecting the communication link (satellite vs aerial). In particular, the shape of the \ac{IQ} clouds at the receiver when considering benign and malicious signals is different--- the signal received from the spoofer installed on the drone has more statistical variance around the expected values than the signal received from the satellites. We attribute this phenomenon to the different fading phenomena affecting the received signals, as the signal generated by the satellites comes from a different altitude than the one originated by the spoofer on the drone, being subject to different fading.

\textcolor{black}{According to such considerations, the key design objective of our solution is to learn an efficient, compressed representation (encoding) of the input \ac{IQ} data, for the purpose of feature learning, while being able to reconstruct the original input as accurately as possible from this encoding. Using mathematical notation, considering: (i) the input $\mathbf{x}$ (image generated from $N$ \ac{IQ} samples), (ii) the encoder function $f_\theta$, with $\theta$ being a parameter that maps $\mathbf{x}$ over the latent representation $\mathbf{z}$, (iii) the decoder function $g(\phi)$, being $\phi$ a parameter that is used for decoding $\mathbf{z}$ into a reconstruction of the input $\mathbf{\tilde{x}}$, and (iv) $\hat{\mathbf{x}} = g_\phi(f_\theta(\mathbf{x}))$ the reconstruction of the input, the objective of our autoencoder is to minimize the reconstruction loss as defined in~\cite{goodfellow2016_book} over the dataset of the training images $D$, according to Eq.~\ref{eq:key_obj}:
\begin{equation}
    \label{eq:key_obj}
    \min_{\theta, \phi} \ \mathbb{E}_{\mathbf{x} \sim \mathcal{D}} \left[ \mathcal{L}(\mathbf{x}, \hat{\mathbf{x}}) \right] = \min_{\theta, \phi} \ \mathbb{E}_{\mathbf{x} \sim \mathcal{D}} \left[ \mathcal{L}(\mathbf{x}, g_\phi(f_\theta(\mathbf{x}))) \right] \|_2^2,
\end{equation}
where $\mathcal{L}$ is the reconstruction loss function introduced in Eq.~\ref{eq:mse_loss}.
}

\begin{figure}
    \centering
    \includegraphics[width=1.0\columnwidth]{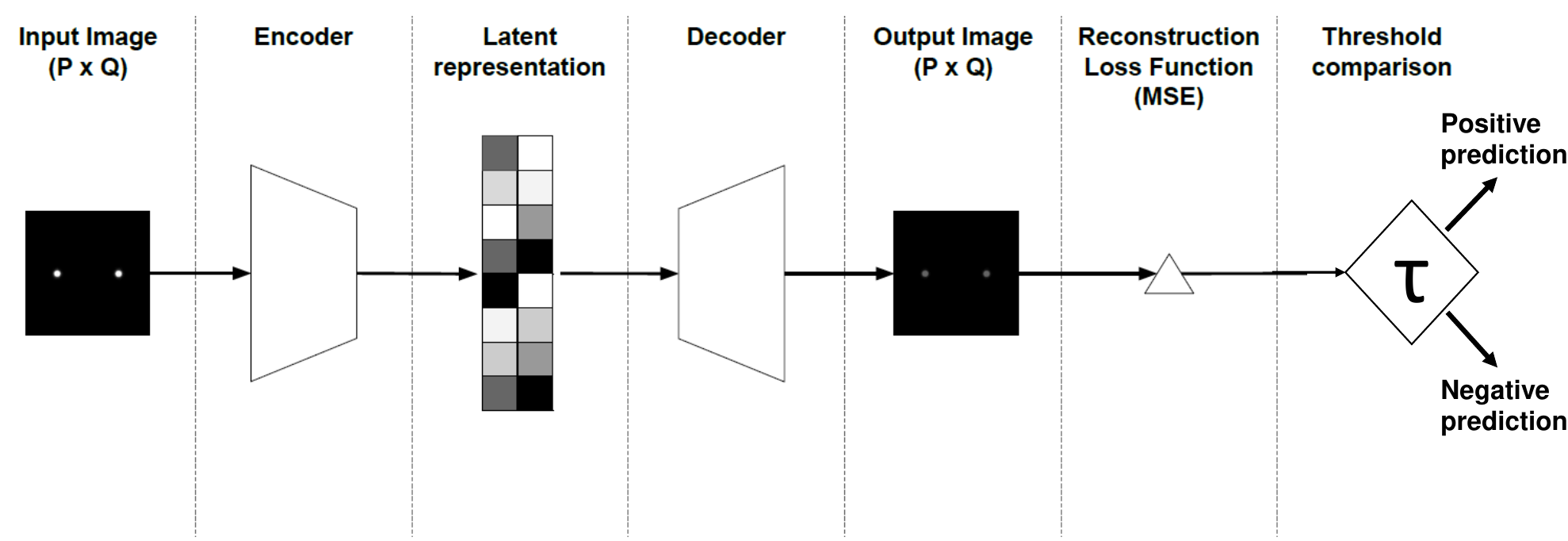}
    \caption{\ac{AE}-based spoofing detection architecture. 
    }
    \label{fig:ae_spoof}
\end{figure}

Figure~\ref{fig:ae_spoof} shows the architecture of the adopted \ac{AE}.
\textcolor{black}{We provide images (generated through the procedure described in Sec.~\ref{sec:image_generation}) as input to an encoder using a logarithmic sigmoid activation function with $L=16$ neurons. Such an operation generates a compressed latent representation of the input image, characterized by $L=16$ dimensions. Specifically, our encoder generates the latent representation $\mathbf{z}$ according to Eq.~\ref{eq:encoder}:
\begin{equation}
    \label{eq:encoder}
    \mathbf{z} = \sigma\left(W^{(1)} \mathbf{x} + \mathbf{b}^{(1)}\right),
\end{equation}
where $( W^{(1)} \in \mathbb{R}^{16 \times n})$ is the encoder weight matrix, $( \mathbf{b}^{(1)} \in \mathbb{R}^{16} )$ is the bias vector of the encoder and $\sigma(s) = \frac{1}{1 + e^{-s}}$ is the sigmoid activation function.
We feed such a latent representation vector $\mathbf{z}$ to a decoder using a linear decoder transfer function, using $S = 50,176$ neurons. Our \ac{AE} architecture uses two hidden layers, the optimizer algorithm Limited-memory Broyden–Fletcher–Goldfarb–Shanno (L-BFGS) algorithm, and the sparsity regularization technique. Thus, the operations performed by the decoder to obtain the reconstructed input $\hat{\mathbf{x}}$ can be expressed as in Eq.~\ref{eq:decoder}.
\begin{align}
    \label{eq:decoder}
    \mathbf{z}^{(2)} &= \phi^{(2)}\left(W^{(2)} \mathbf{z}^{(1)} + \mathbf{b}^{(2)}\right), \quad \mathbf{z}^{(2)} \in \mathbb{R}^{h_2} \\ \nonumber
    \mathbf{z}^{(3)} &= \phi^{(3)}\left(W^{(3)} \mathbf{z}^{(2)} + \mathbf{b}^{(3)}\right), \quad \mathbf{z}^{(3)} \in \mathbb{R}^{h_3} \\ \nonumber
    \hat{\mathbf{x}} &= W^{(4)} \mathbf{z}^{(3)} + \mathbf{b}^{(4)}, \quad \hat{\mathbf{x}} \in \mathbb{R}^{50176}
\end{align}
where $( \phi^{(2)}, \phi^{(3)} )$ are activation functions for decoder hidden layers, $( W^{(2)}, W^{(3)}, W^{(4)} )$ are weights for each decoder layer, $( \mathbf{b}^{(2)}, \mathbf{b}^{(3)} )$, are biases for each decoder layer, and $\mathbf{z}^{(2)}$ and $\mathbf{z}^{(3)}$ represent the intermediate activations (or outputs) of each decoder hidden layer, namely the hidden states. Note that, in line with general principles of autoencoders (see~\cite{goodfellow2016_book}), for a deep network with $L$ layers there are $L+1$ weight matrices, i.e., one for each layer and one final weight matrix to project from the last hidden layer to the output. We obtain as output $\hat{\mathbf{x}}$ an image of the same size as the input ($P\times Q$), which we convert into a matrix and compare with the input image through the \ac{MSE} loss function, as in Eq.~\ref{eq:mse_loss} (Sec.~\ref{sec:background}). 
} 

Our approach involves \emph{training} and \emph{testing}. For the training process, we use only images generated from \ac{IQ} samples captured from legitimate LEO satellite communications, generating a set of \ac{MSE} values characterizing the regular pattern of the communication link, namely $\mathbf{MSE_{train}}$. We use such values to compute the threshold $\tau$, used to differentiate images generated from legitimate satellites from spoofed ones. An image with $MSE \le \tau$ is predicted to be generated from legitimate transmitters (LEO satellites), while an image with $MSE > \tau$ is predicted to be generated from unauthorized entities (adversary). For computing $\tau$, we use Eq.~\ref{eq:Threshold_Formula}, as proposed by the authors in \cite{erfani2016_prec} (Sec.4.1), so as to take into account as much as possible the variability of legitimate communication patterns due to noise.
\begin{equation}
    \label{eq:Threshold_Formula}
    \tau = E(\mathbf{MSE_{train}}) + 3 \cdot \sigma(\mathbf{MSE_{train}}),
\end{equation} 
where $E(\circ)$ and $\sigma(\circ)$ represent the mean and the standard deviation, respectively.

\begin{figure}
    \centering
    \includegraphics[width=\columnwidth]{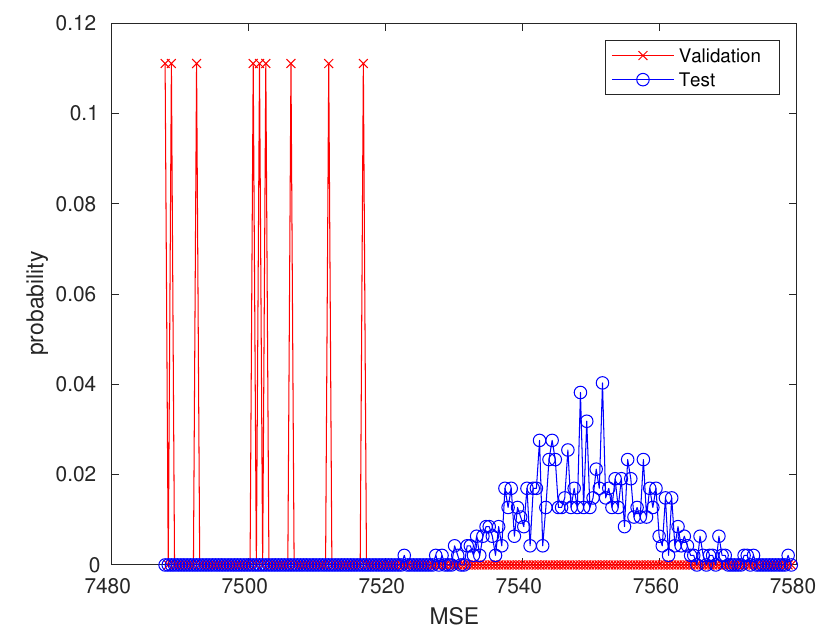}
    \caption{Distribution of \ac{MSE} values from our dataset: red crosses represent \ac{MSE} computed on satellite images, and blue circles represent \ac{MSE} values computed on images generated from the aerial link (drone).
    }
    \label{fig:MSE_Distr}
\end{figure}

An image generated from satellite samples is characterized by lower \ac{MSE} values. In contrast, an image generated from a transmitter at a lower altitude (terrestrial or aerial) results in a higher \ac{MSE} value. 
We show a sample distribution of the \acp{MSE} in Fig.~\ref{fig:MSE_Distr}. Note that images generated from satellite data are characterized by a higher probability of having lower \ac{MSE} values compared to images generated from data spoofed by the drone. Thus, by considering a discrimination threshold, we can detect spoofing attacks as anomalies in the resulting \ac{MSE} values.
Note that we empirically select a hidden size value of $16$, a sparsity regularization term of $0.5$, an L2-regularization term of $0.01$, and $250$ epochs for the training. Such values were selected through experimental fine-tuning based on the available data (see later section).

\section{Data Collection}
\label{sec:data_collection}

In this section, we describe the experimental setup used for data collection (Sec.~\ref{sec:experimental_setup}), the experiment settings (Sec.~\ref{sec:exp_settings}), and we report on a preliminary statistical analysis of our measurements (Sec.~\ref{sec:data_analysis}).

\subsection{Experimental Setup}
\label{sec:experimental_setup}

\begin{figure*}[h]
    \centering
    \begin{subfigure}[b]{.67\columnwidth}
        \centering
        \includegraphics[width=.98\columnwidth]{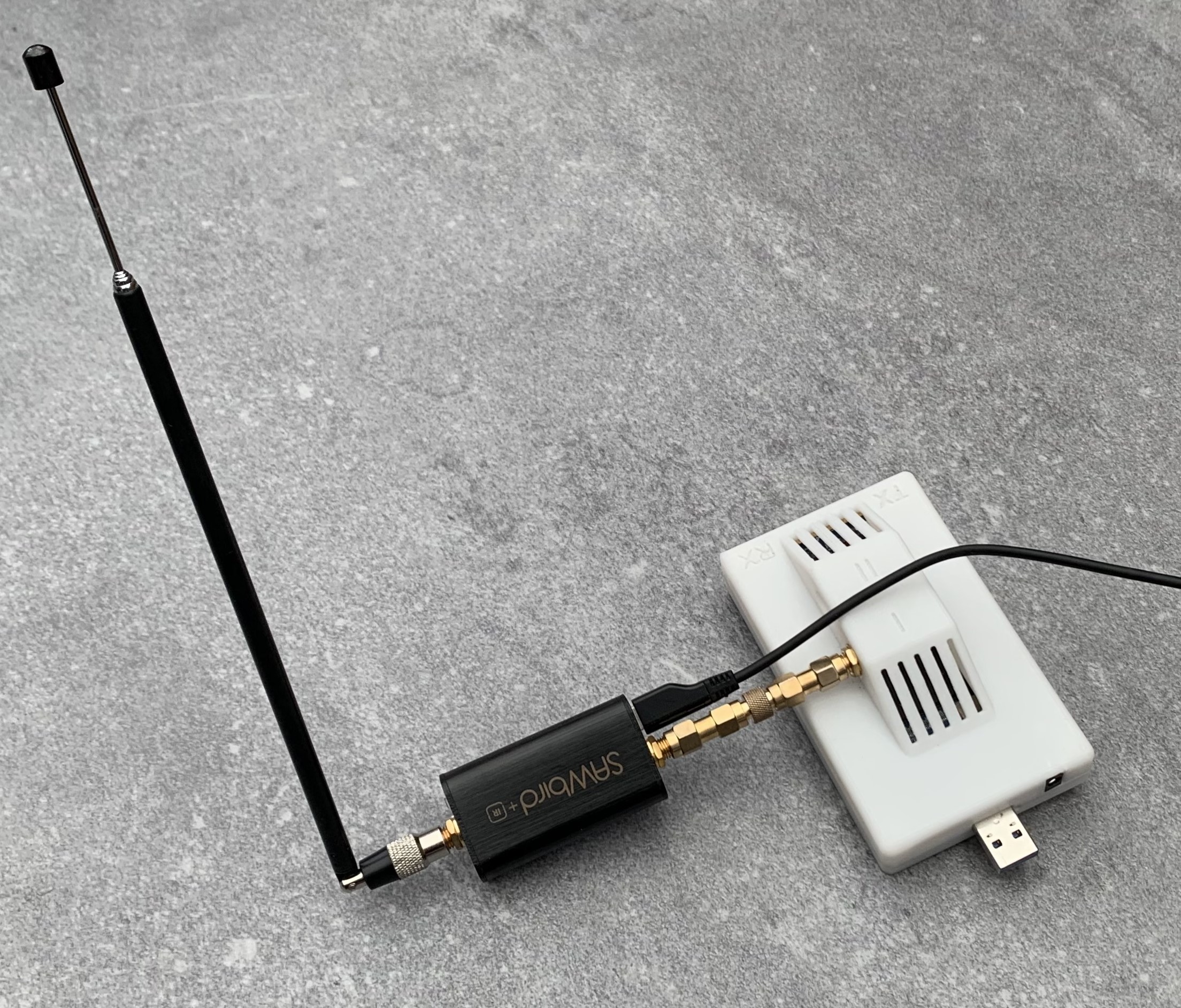}
        \caption{Receiver setup.}
        \label{fig:rx_setup}
    \end{subfigure}
    \begin{subfigure}[b]{.67\columnwidth}
        \centering
        \includegraphics[width=.9\columnwidth]{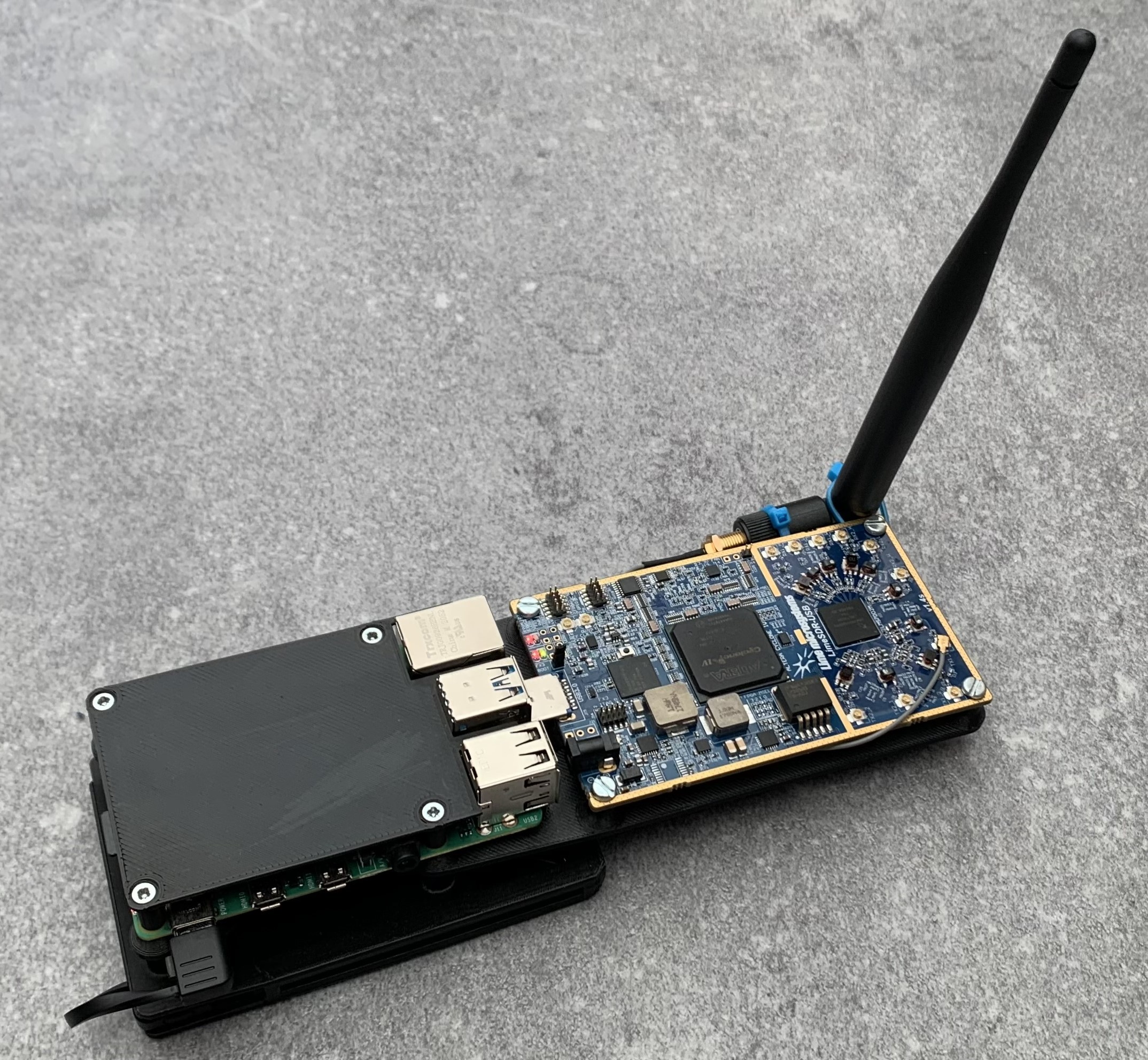}
        \caption{Spoofing setup}
        \label{fig:tx_setup}
    \end{subfigure}
    \begin{subfigure}[b]{.67\columnwidth}
        \centering
        \includegraphics[width=\columnwidth]{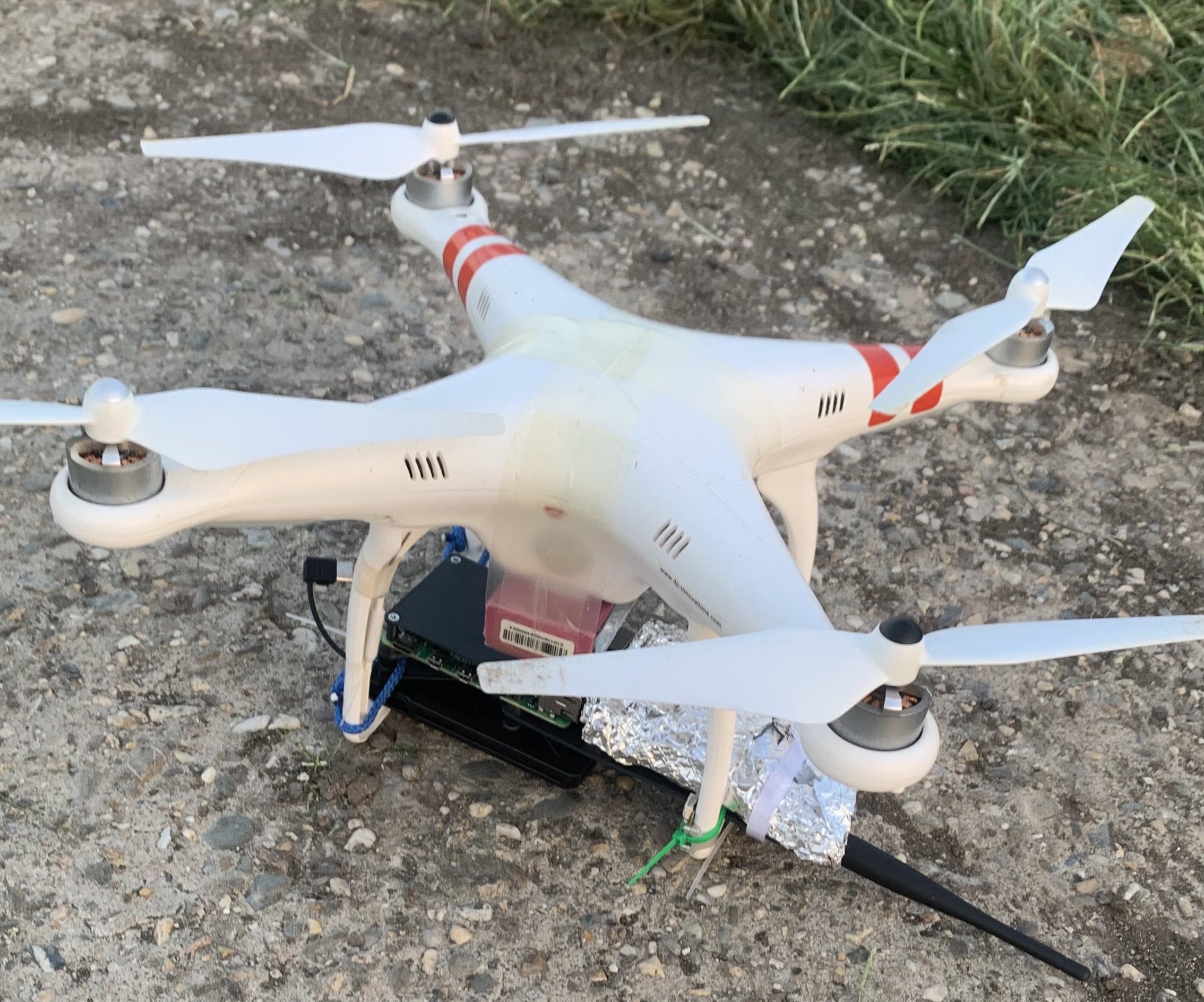}
        \caption{Spoofing setup mounted on the drone.}
        \label{fig:tx_drone_setup}
    \end{subfigure}
    \caption{Experimental setup, including the receiver setup (a), spoofing setup (b) and spoofer installation on the drone (c).}
    \label{fig:setup}
\end{figure*}

{\bf Hardware Details}. To collect the data required to test our solution, we use two \acp{SDR} \emph{LimeSDR USB}. This hardware has a frequency range of $100$~kHz up to $3.8$~GHz, a maximum bandwidth of $61.44$~MHz, and a transmit power of up to $10$~dBm. One LimeSDR is used as a receiver of the IQ samples (the user in our scenario). To boost message reception, we use the \ac{LNA} module \textit{Nooelec SAWbird+IR}, characterized by high attenuation outside of the 60MHz bandpass region centered around 1.620GHz, and a minimum of 20dB of gain within the bandpass region, being suitable for receiving messages in the bandwidth of our interest~\cite{nooelec}. 
We connect the receiving LimeSDR and the LNA to a telescopic antenna, extendable on purpose to maximize access to the open sky. Finally, we connect the LimeSDR USB to a laptop, running software used to communicate with the SDR. Figure~\ref{fig:setup}(a) illustrates the overall receiver (RX) setup.
Another LimeSDR is used as a transmitter (spoofer) of IRIDIUM messages and installed onboard a drone. We use the drone DJI Phantom 1, characterized by a maximum vertical speed of $6$~m/s, maximum horizontal speed of $10$~m/s, and a flight time of approx. $15$~minutes. Onboard the drone, besides the LimeSDR (weighting $49$~g), we install the embedded computer Raspberry Pi 4, allowing us to control the LimeSDR and weighting $87$~g, and a power bank \emph{TNTOR 3500mAh}, weighting $94$~g and providing enough input power to the Raspberry Pi (see Fig.~\ref{fig:setup}(b)). To hold all such components together in a safe way, we use 3D-printed parts weighing approx. $20$~g, for a total payload weight of $250$~g that can be carried by the selected drone with no impact on stability and maneuverability~\cite{DroneCarry},~\cite{droneblog_payload}. With such a weight onboard, the drone battery lasts approx. 10 minutes at $120$~m, allowing us to gather enough data for our experiments. 
We show the overall transmitter setup in Fig.~\ref{fig:setup}(c).\\

\noindent
{\bf Software Details.} On the laptop connected to the receiver, we use the state-of-the-art software \emph{iridium-extractor} for receiving and decoding IRIDIUM messages~\cite{gr_iridium}.  
We modified the tool to output raw IQ samples of decoded packets besides the demodulated bits of the messages.
To transmit valid IRIDIUM messages, we first collected locally a dataset of legitimate IRIDIUM packets, and then used the LimeSDR installed on the drone to replay such packets on the same carrier frequency used by legitimate IRIDIUM communications, through the same modulation (DQPSK). To do this, we created a custom flow chart using GNURadio, a popular software development toolkit for \acp{SDR}, and we verified that our injected IRIDIUM messages are received and processed with no errors by \mbox{\emph{gr-iridium}}, as legitimate messages. To enable message injection on the LimeSDR installed on the drone, we connect via SSH to the Raspberry Pi on a wireless connection through our laptop, and we use a Tmux session to allow our program to run also when the SSH session is closed.

\subsection{Experiment Settings}
\label{sec:exp_settings}
\textcolor{black}{
We deployed the described experimental setup at a rural field located about $1.5$~km northeast of Herwijnen, in the Netherlands, shown in Fig.~\ref{fig:data_aq_field}.
\begin{figure}[h]
    \centering
    \includegraphics[width=\columnwidth]{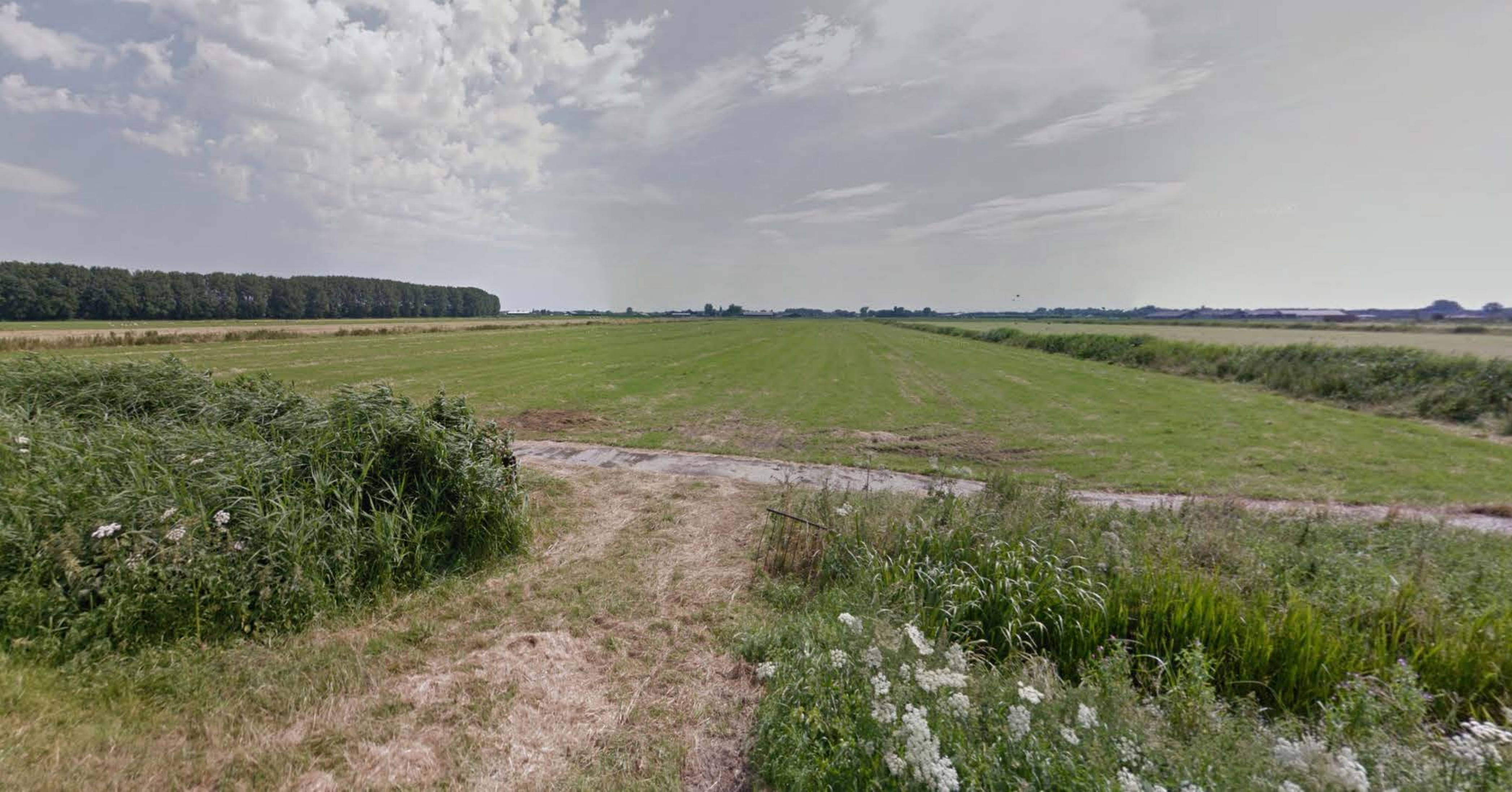}
    \caption{Photo of the data collection site. }
    \label{fig:data_aq_field}
\end{figure}
The field is located at a minimum distance of approx. $500$~m from any man-made construction, offering an ideal place to fly the drone while minimizing safety concerns. Moreover, for reproducibility purposes, note that there is only grass at the ground level that impacts the profile of the signal received from satellites and drones.
At this location, we acquired legitimate and spoofed data while considering the following three scenarios:
\begin{itemize} 
    \item \emph{S1: Outdoor - Static.} We consider the receiver stationary on the ground and the transmitter drone hovering in the air, at increasing altitude up to $120$~m, i.e., the maximum allowed altitude for a drone in our area~\cite{ilent_nl}.
    \item \emph{S2: Outdoor - Moving Transmitter.}  We consider the receiver stationary on the ground and the transmitter on the drone moving in the air at increasing altitude up to $120$~m, in a circle of radius $\approx 15$~m around the receiver. We flew this movement manually by moving the sticks on the remote controller. This scenario aims to estimate the detection performance when the adversary implements evasion techniques by moving itself (flying around).
    \item \emph{S3: Outdoor - Moving Receiver and Transmitter.} We consider both the receiver and the transmitter moving in a circle of radius $\approx 15$~m from the starting location, coincidental with the receiver location in S1 and S2. We let the drone fly at an increasing altitude, up to $120$~m. During the experiments, the transmitter and receiver experience various (random) distances. This scenario aims to investigate the impact of receiver movement on the capability to detect spoofing.
\end{itemize}
We believe that such scenarios cover the most realistic cases of aerial spoofing attacks to LEO satellite receivers deployed outdoors, making our results generalizable to any receivers deployed outdoors in a scenario like the one shown in Fig.~\ref{fig:data_aq_field}. Moreover, our experiments consider various movement patterns, accounting for heterogeneous noise patterns, multipath, and interference conditions characterizing real-world satellite communication channels. Our experiments also cover multiple altitudes and fading dynamics, addressing possible concerns about the limited scalability and robustness of our experimental findings. Note that, to avoid interfering with other potential IRIDIUM receivers in the area, we selected a TX power gain value on the transmitter low enough not to generate interference further away than $500$~m, while being high enough to still receive messages from the maximum altitude of $120$~m. Finally, note that we did not perform new measurements for ground-based spoofing attacks, as such scenarios are already covered by the literature~\cite{oligeri2024sac}.
}

\subsection{Data Analysis}
\label{sec:data_analysis}

\begin{figure}[h]
\centering
    \subfloat[Legitimate IQ Samples]{\includegraphics[width=0.49\columnwidth]{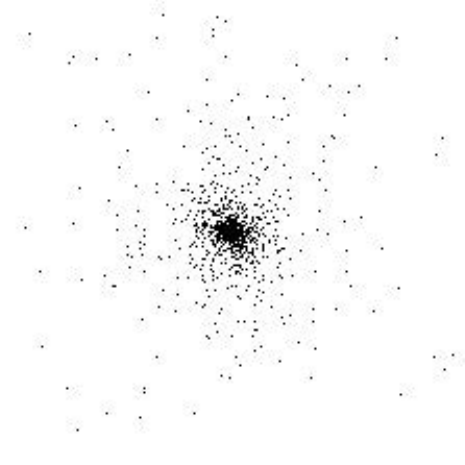}}
    \hfill
    \subfloat[Spoofed IQ Samples]{\includegraphics[width=0.49\columnwidth]{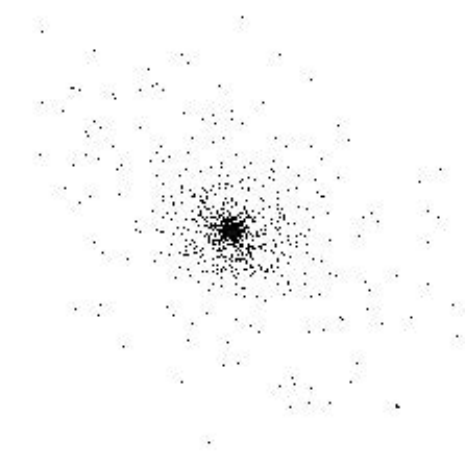}}
    \caption{$N=1,000$ I-Q samples as received from legitimate satellites (a) and drones (b).
    }
    \label{fig:images}
\end{figure}

\begin{figure}[h!]
    \centering
    \includegraphics[width=\columnwidth]{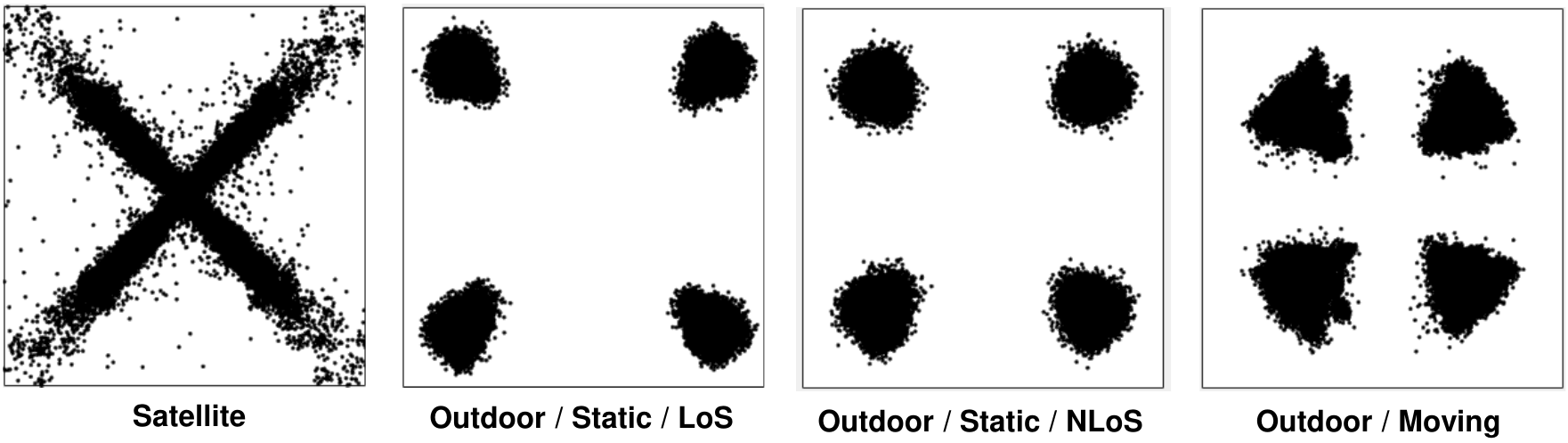}
    \caption{IQ samples collected as part of the analysis in~\cite{oligeri2024sac}, received on an Ettus X310 from (i) Satellite, (ii) Spoofer deployed Outdoors, with Static LoS, (iii) Spoofer deployed Outdoors, with Static NLoS, and (iv) Spoofer deployed Outdoors and moving. Each sub-figure reports a subset of 10,000 I-Q samples randomly taken from the collected measurements.
    }
    \label{fig:iq_satprint}
\end{figure}

\begin{figure}[h!] 
    \centering
    \begin{subfigure}[b]{0.49\columnwidth}
        \centering
        \includegraphics[width=\columnwidth]{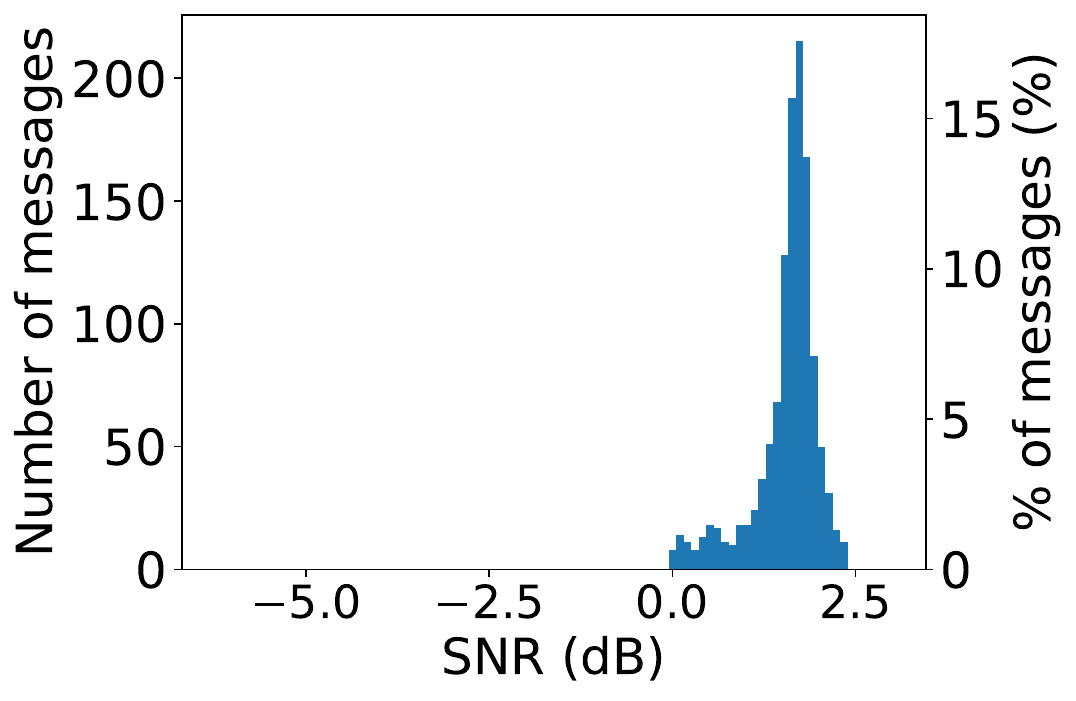}
        \caption{Legit. msgs. - $N=250$.}
        \label{fig:snr_values_benign_250}
    \end{subfigure}
    \begin{subfigure}[b]{0.49\columnwidth}
        \centering
        \includegraphics[width=\columnwidth]{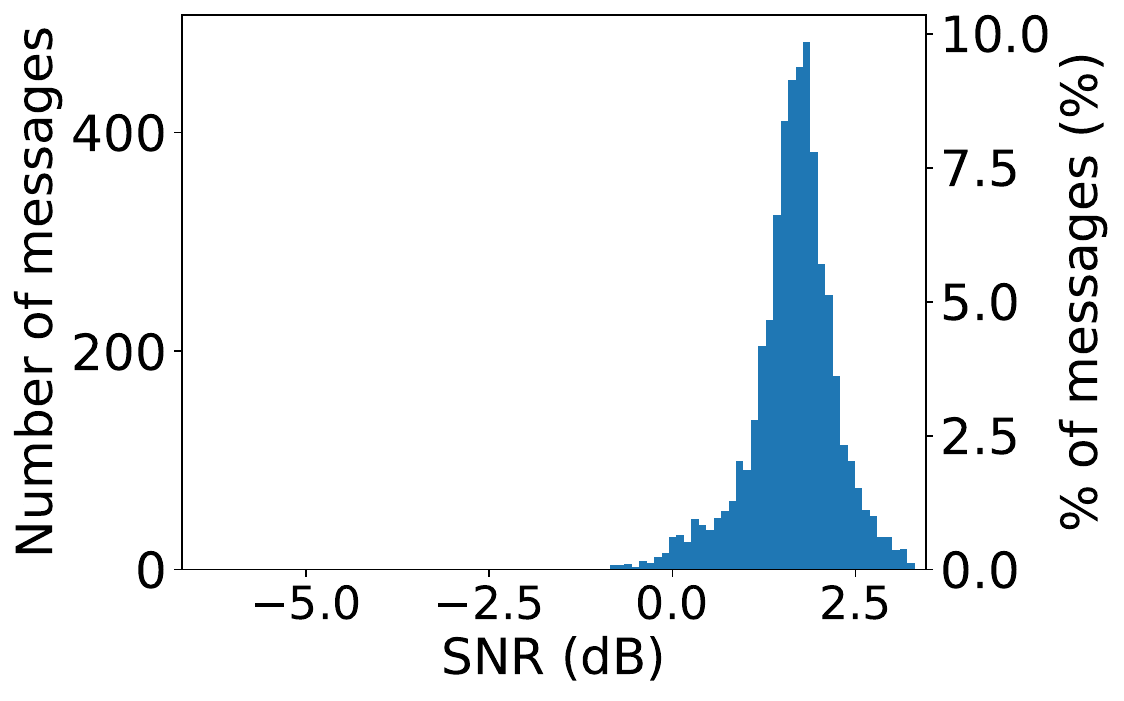}
        \caption{Legit. msgs., $N=1,000$.}
        \label{fig:snr_values_benign_1000}
    \end{subfigure}
    \begin{subfigure}[b]{0.49\columnwidth}
        \centering
        \includegraphics[width=\columnwidth]{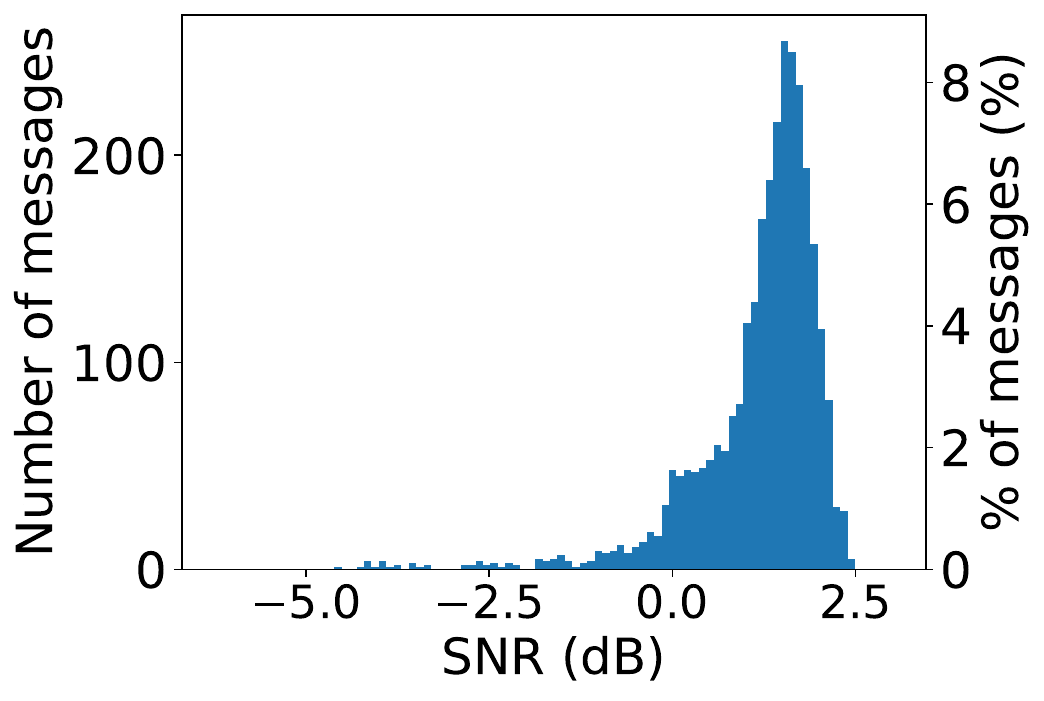}
        \caption{Spoofed msgs. - $N=250$.}
        \label{fig:snr_values_spoofed_250}
    \end{subfigure}
    \begin{subfigure}[b]{0.49\columnwidth}
        \centering
        \includegraphics[width=\columnwidth]{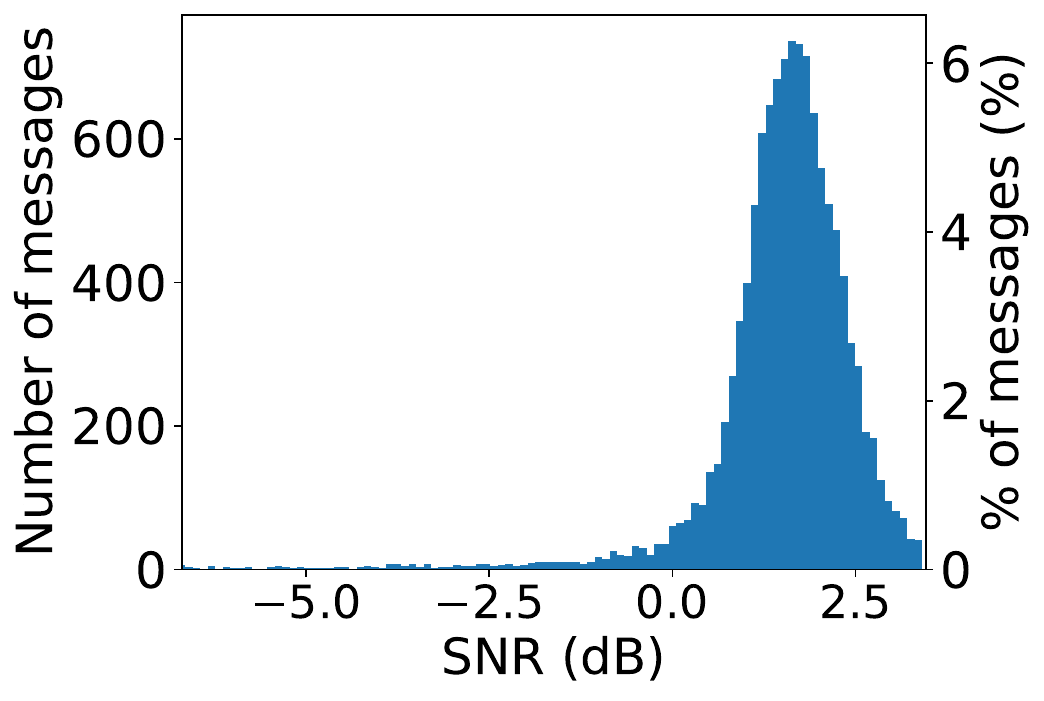}
        \caption{Spoofed msgs. - $N=1,000$.}
        \label{fig:snr_values_spoofed_1000}
    \end{subfigure}
    \caption{Distribution of the \ac{SNR} of received legitimate (a, b) and spoofed (c, d) IRIDIUM messages. 
    }
    \label{fig:snr_values}
\end{figure}

Our collected dataset includes more than $10,000$ legitimate IRIDIUM messages. Moreover, we collected more than $1,000$ spoofed IRIDIUM messages for each scenario and tested altitude, as summarized in Tab.~\ref{tab:messages}, for a total number of approx. $24,000$ messages. Our dataset is available as open source at~\cite{dataset}.
\begin{table}[h]
\centering
\begin{tabular}{|c|cc|}
\hline
\multicolumn{1}{|l|}{{\bf Experiment}} & {\bf  Altitude [m]} &  {\bf Messages [\#]} \\ \hline
\multirow{6}{*}{S1}            & \multicolumn{1}{c}{10} &    1110             \\
                               & 30                     &    1008             \\
                               & 50                     &    1163            \\
                               & 70                     &    1133            \\
                               & 90                     &    1263             \\
                               & 120                    &    1104 \\ \hline   
\multirow{3}{*}{S2}            & \multicolumn{1}{c}{30} &    1189      \\
                               & 70 & 1178 \\
                               & 120 & 1238 \\ \hline
\multirow{3}{*}{S3}            & \multicolumn{1}{c}{30} &    1169      \\
                               & 70 & 1085 \\
                               & 120 & 1239 \\ \hline
\end{tabular}
\caption{Collected spoofed messages.}
\label{tab:messages}
\end{table}
First, note that our IRIDIUM receiver detects all the spoofed messages as legitimate IRIDIUM messages, as they were delivered from IRIDIUM satellites. This successful experiment demonstrates the feasibility of launching (aerial) spoofing attacks to a LEO satellite constellation, i.e., IRIDIUM, using the unencrypted broadcast channel IRIDIUM Ring Alert.
We report in Fig.~\ref{fig:images} example images generated from $N=1,000$ legitimate and spoofed IQ samples, highlighting the similarity of the received data when the user is subject to spoofing attacks compared to a benign scenario. When comparing the images in Fig.~\ref{fig:images} with the images in Fig.~\ref{fig:iq_satprint}, reporting the profile of the received IQ samples collected from ground-based spoofers as part of the analysis carried out in the preliminary contribution in~\cite{oligeri2024sac}, it is immediate to visualize that aerial spoofing attacks are more challenging to detect that ground-based spoofing attacks, motivating our research and contribution.
Figure~\ref{fig:snr_values} also reports the distribution of the \ac{SNR} of legitimate and spoofed messages when $N=1,000$ and $N=250$. We computed the SNR according to the geometrical formula given in Eq.~\ref{eq:snr_final}~\cite{irfan2024_arxiv}.
\begin{equation}
    \label{eq:snr_final}   
    \begin{split}
        SNR & = P(r) - P(n) \\
            & = 10 \cdot \log_{10}{(I^2 + Q^2)} - 10 \cdot \log_{10}{((I-1)^2 + Q^2)} \\
            & = 10 \cdot \log_{10}{\frac{I^2 + Q^2}{(I-1)^2 + Q^2}}.
    \end{split}
\end{equation}
We notice minimal differences in the \ac{SNR} values of legitimate and spoofed signals. Taking into account the images generated from $N=250$ samples, only $7.7$\% of the images generated from the spoofed samples report SNR values outside of the interval of SNR values characterizing images generated from legitimate samples, which is $[0, 2.5]$~dBm. Similarly, when considering images generated from $M=1,000$ samples, the overlap area between the two distributions reduces further to $3.3$\%.
These results indicate that, independently of the number of samples, we cannot classify the images (or the associated \ac{IQ} samples) as legitimate or spoofed by only looking at the \ac{SNR}, suggesting the need for a tailored detection solution.

\section{Experimental Performance Analysis}
\label{sec:performance}

In this section, we report the main results of our experimental performance assessment. We explain the metrics used for our results in Sec.~\ref{sec:metrics} and report our results in Sec.~\ref{sec:results}.

\subsection{Metrics}
\label{sec:metrics}

\textcolor{black}{
In the following, we report the performance of all the tested solutions using the \ac{AUC} of the \ac{ROC} curve. The ROC curve plots the True Positive Rate (TPR) against the False Positive Rate (FPR) at various threshold settings.
Therefore, the ROC curve inherently depends on how false positives (FP) vary with different classification thresholds.
The AUC is the area under this ROC curve, so it summarizes how well the classifier distinguishes between classes across all thresholds, taking into account both true positives and false positives~\cite{bishop2006_book}. This metric ranges between the values $0$ and $1$, with $0.5$ representing a random guess and $1$ indicating a perfect classification. Note that the \ac{ROC} \ac{AUC} measures the quality of a classifier independently of the threshold selection process, and it is recommended when the datasets are unbalanced, as it evaluates the trade-off between sensitivity (true positive rate) and specificity (true negative rate) in various threshold settings~\cite{khan2017_esa},~\cite{angiulli2024_arxiv}. 
}

For each reported figure, we train our proposed solution as outlined in Sec.~\ref{sec:methodology}, using $80$\% of the images generated from legitimate samples. We validate the model using the remaining $20$\% and then test it using spoofed data acquired in a particular scenario, as described in Sec.~\ref{sec:data_collection}.
We use K-fold validation with $K=5$ for each configuration and report the quantiles 0.05, 0.5, and 0.95 for the ROC AUC values.
\textcolor{black}{Finally, we compare our proposed solution against two benchmarks, i.e., the solution in~\cite{oligeri2024sac}, using a binary \ac{FCDD} to detect spoofing attacks~\cite{liznerski2021explainable}, and the proposal in~\cite{oligeri2023tifs}, using the neural network \emph{Resnet-18} to authenticate satellite wireless messages. Note that the proposal in~\cite{oligeri2024sac} considers the same problem tackled in this manuscript. Thus, no changes are needed to ensure a fair comparison. Conversely, to allow for a fair comparison between our setup and the proposal in~\cite{oligeri2023tifs}, we use the neural network classifier \emph{Resnet-18} to distinguish between satellite (authorized) and drone (unauthorized) transmissions using binary classification.}

We ran all data analyses on the \ac{HPC} cluster provided by TU/e, Eindhoven, Netherlands, including a CPU AMD EPYC 7313 running at 3.00 GHz, 16 GB of RAM, and a TESLA V100 GPU with 16GB of RAM.

\subsection{Results}
\label{sec:results}

\begin{figure*}[h]
    \centering
    \begin{subfigure}[b]{0.67\columnwidth}
        \centering
        \includegraphics[width=\columnwidth]{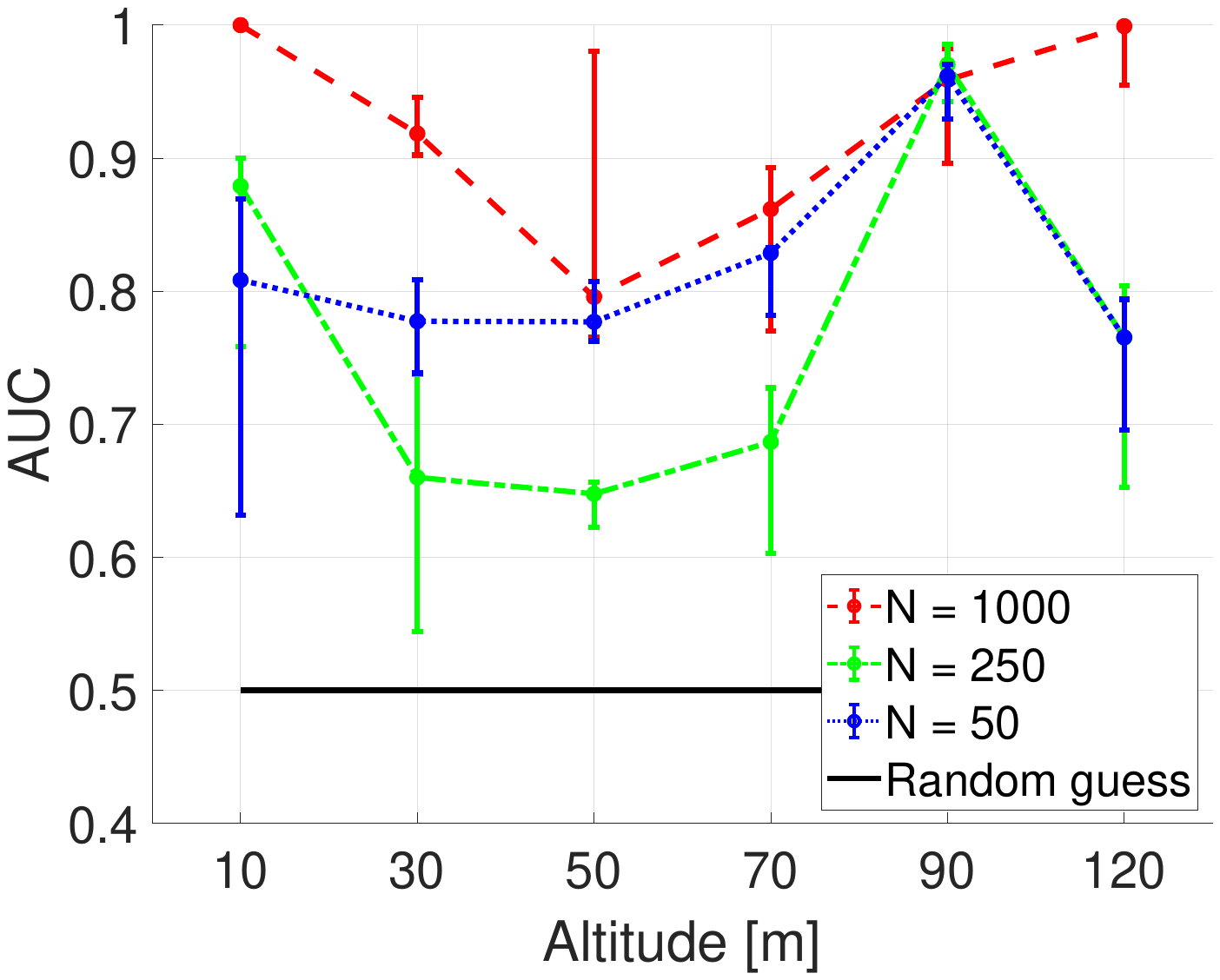}
        \caption{S1.}
        \label{fig:comp_e1}
    \end{subfigure}
    \begin{subfigure}[b]{0.67\columnwidth}
        \centering
        \includegraphics[width=\columnwidth]{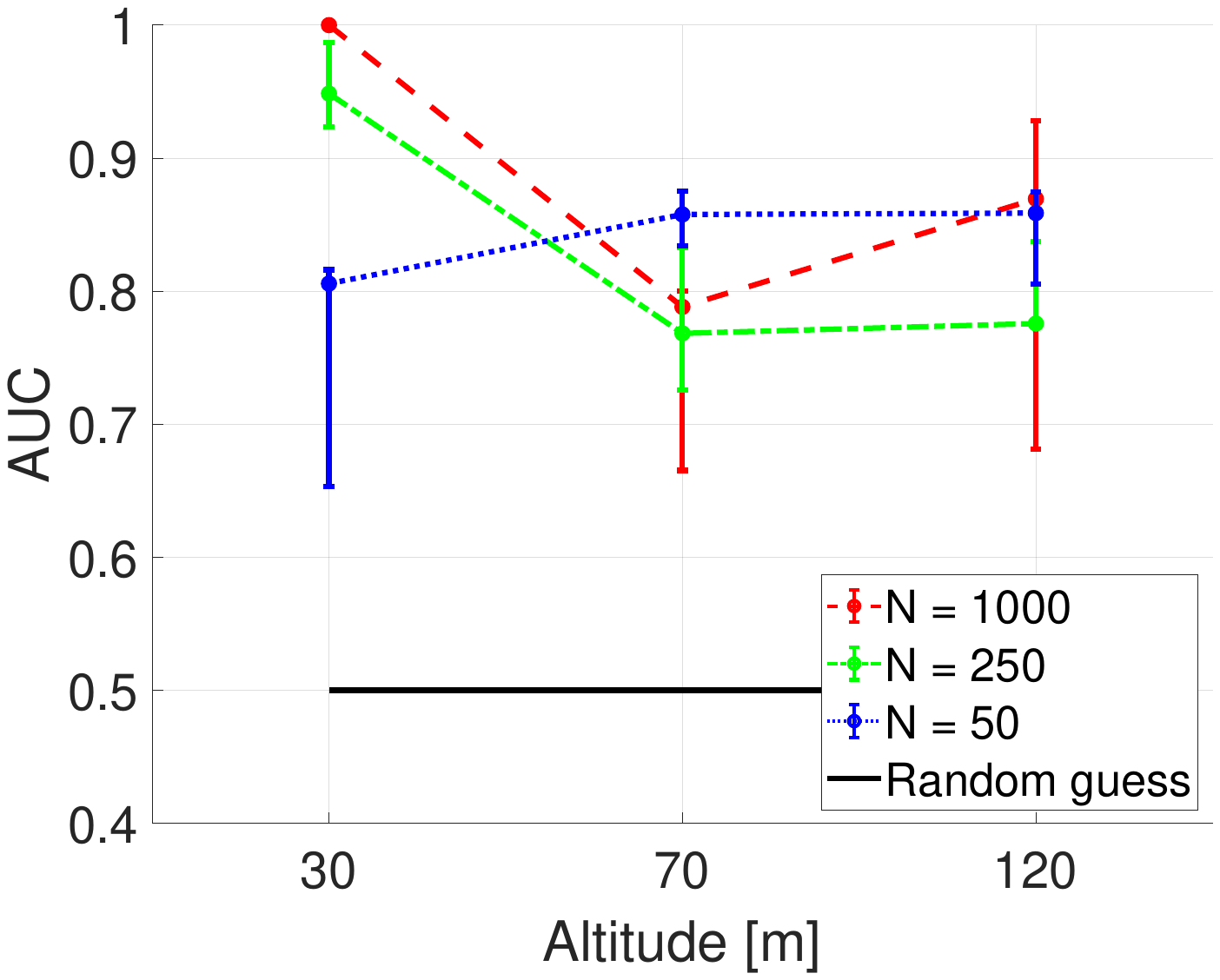}
        \caption{S2.}
        \label{fig:comp_e2}
    \end{subfigure}
    \begin{subfigure}[b]{0.67\columnwidth}
        \centering
        \includegraphics[width=\columnwidth]{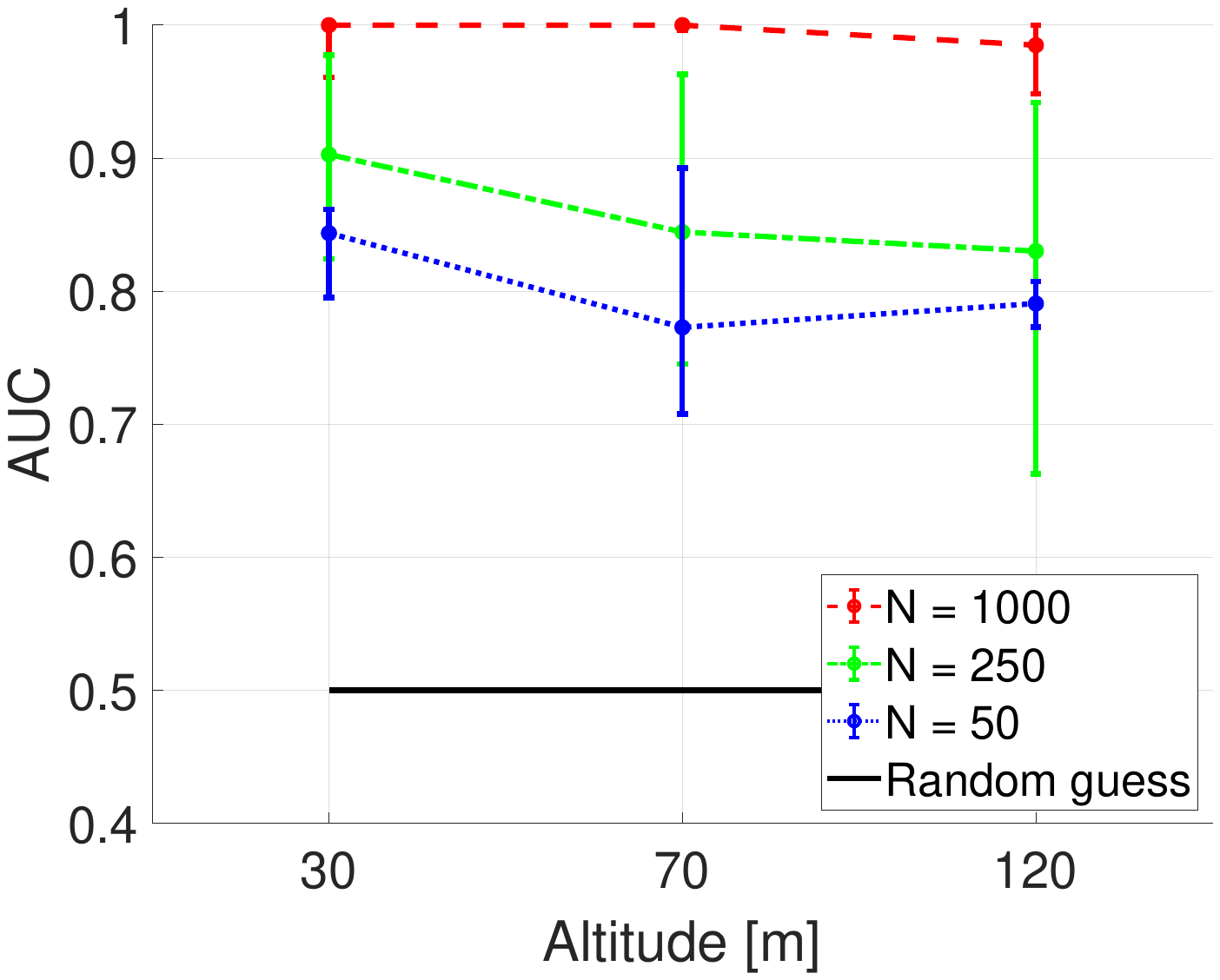}
        \caption{S3.}
        \label{fig:comp_e3}
    \end{subfigure}
    \caption{Performance of our solution in S1 (a), S2 (b), and S3 (c), with $N = 50$, $N = 250$, and $N=1,000$ samples per image.}
    \label{fig:results_autoencoder}
\end{figure*}

We report in Fig.~\ref{fig:results_autoencoder} (a-c) the performance of our solution in the three investigated scenarios, using various numbers of samples per image $N$ (recall Sec.~\ref{sec:image_generation}). 
In the vast majority of cases, the protocol configuration using $N=1,000$ samples per image outperforms the configurations using fewer samples, demonstrating that, generally speaking, using a higher number of samples per image increases spoofing detection robustness. This finding is likely due to the fact that using more samples per image helps to better isolate outlier samples. When histograms are not used for generating images, such outliers have the same relative weight in the image as the other samples, thus negatively influencing the detection process. We notice just a few cases where using fewer samples per image achieves higher values of the ROC \ac{AUC}, e.g., S1 at $90$~m and S2 at $70$~m. In these few cases, we notice more noise affecting the collected data; thus, using limited samples per image with noisy data can improve the performance by reducing the impact of the noise. Overall, with $N=1,000$ samples per image, our solution can detect spoofing attacks with remarkable accuracy, i.e., with an average ROC AUC always higher than $0.8$ in S1, $0.75$ in S2 and $0.99$ in S3. Considering that IRIDIUM satellites transmit at $50$~kbps, we need $20$~ms to acquire such a number of samples, while running the \ac{AE} requires only a few ms. In all scenarios, we also see a general decreasing trend in the performance of our solution with the increase of the altitude of the spoofing drone. Such results confirm that, to increase the chances of bypassing our solution, the spoofing drone should exceed the maximum height imposed by law regulations and get as high as possible in the sky, closer to the satellites, to imitate the (satellite) fading characteristics more effectively. Besides being against the law, such requirement also forces the attacker to use more powerful and expensive drones, capable of flying at very high altitudes. We also highlight that, as shown by the results in Fig.~\ref{fig:results_autoencoder}(c), the receiver movement introduced in S3 improves performance compared to other scenarios, with any value of $N$. This result is particularly promising for the application of our solution on mobile entities, e.g., vehicles or drones, which already move as part of their regular operations.\\
\textcolor{black}{We also measured the overhead of our solution, i.e., the time required to generate an image from incoming \ac{IQ} samples, namely $\tau_{img}$, and the time to classify such image as benign or spoofed, namely $\tau_{ae}$. We also estimated the memory requirements of our solution, i.e., the size on disk of the model ($M_{ae}$) and the size of the image on disk with various no. of samples $M_{image}$. We report the results in Tab.~\ref{tab:overhead}.
\begin{table}[!h]
\footnotesize
\centering
\begin{tabular}{|l|c|c|l|}
\hline
Samples per Image (N) & \multicolumn{1}{c|}{$250$} & \multicolumn{1}{c|}{$500$} & $1000$ \\ \hline
$\tau_{img}$ (ms) & 0.0085 & 0.0089 & 0.0102 \\ \hline
$\tau_{ae}$ (ms)  & 0.0027 & 0.0028 & 0.0032 \\ \hline
$M_{ae}$ (MB)     & 13     & 13     & 13   \\ \hline
$M_{image}$ (KB)  & 3.6    & 4.35   & 5.76 \\ \hline 
\end{tabular}
\caption{Overhead components of our solution.}
\label{tab:overhead}
\end{table}
We notice that collecting samples necessary to generate an image with $N=250$, $N=500$, or $N=1,000$ samples is in the order of a few micro-seconds ($10$~$\mu$sec with $N=1,000$ samples). Even in the worst case where we use $N=1,000$, the time to classify such an image as benign or malicious on our computing equipment is very limited, i.e., $0.032$~msec, and the generated model requires only $13$~MB of storage space. Therefore, the computational cost of the solution is very limited, as well as its storage requirements are limited for regular computing units. 
}
\textcolor{black}{We further consider the configuration of our solution using $N=250$ samples per image (often performing the worst in Fig.~\ref{fig:results_autoencoder}), and we also compare the performance of our proposed solution to the benchmarks proposed by Oligeri et al. in~\cite{oligeri2024sac} and~\cite{oligeri2023tifs}. We report the results of our comparison in Figs.~\ref{fig:comparison}(a-c).
\begin{figure*}[h]
    \centering
    \begin{subfigure}[b]{0.67\columnwidth}
        \centering
        \includegraphics[width=\columnwidth]{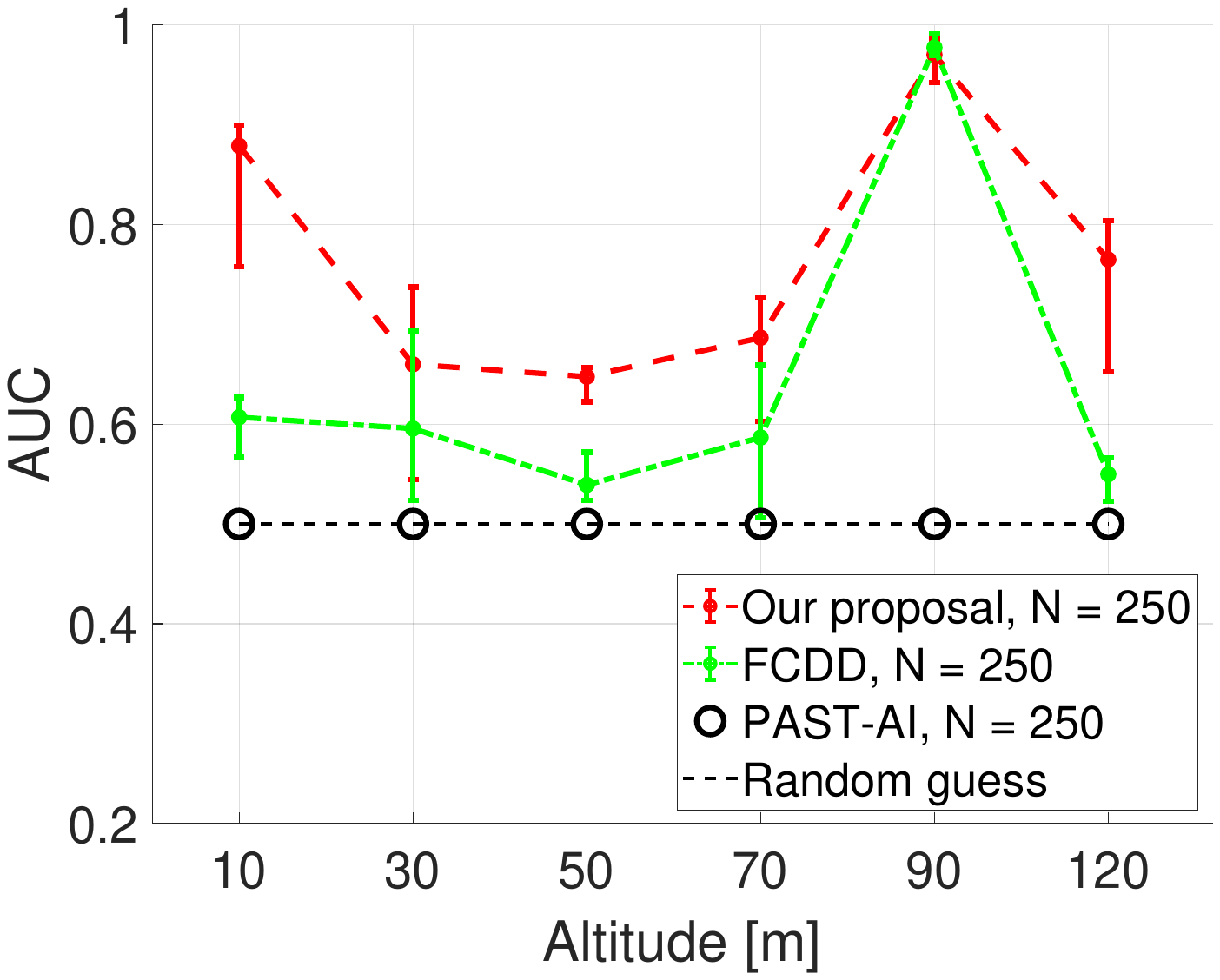}
        \caption{S1.}
        \label{fig:results_e1}
    \end{subfigure}
    \begin{subfigure}[b]{0.67\columnwidth}
        \centering
        \includegraphics[width=\columnwidth]{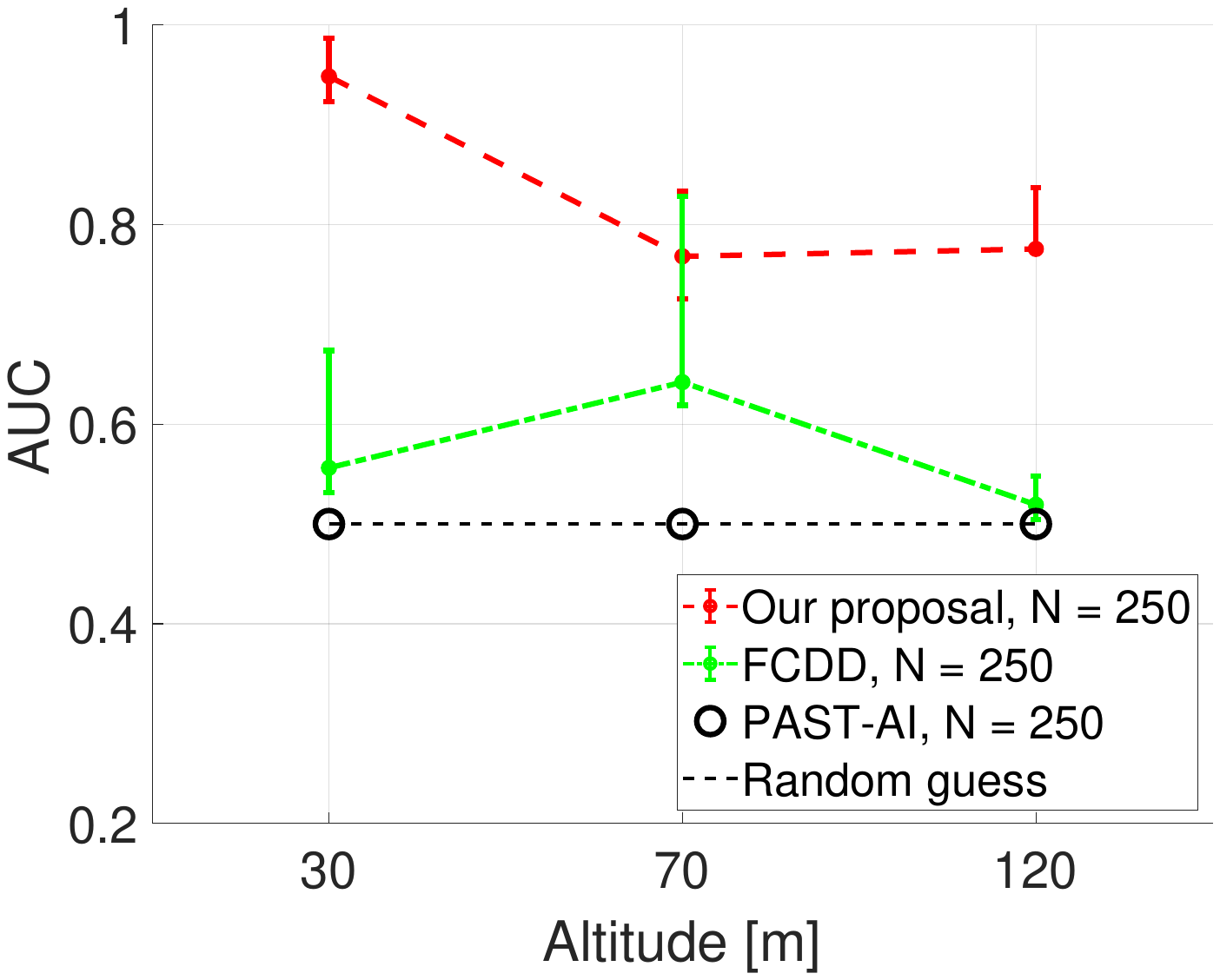}
        \caption{S2.}
        \label{fig:results_e2}
    \end{subfigure}
    \begin{subfigure}[b]{0.67\columnwidth}
        \centering
        \includegraphics[width=\columnwidth]{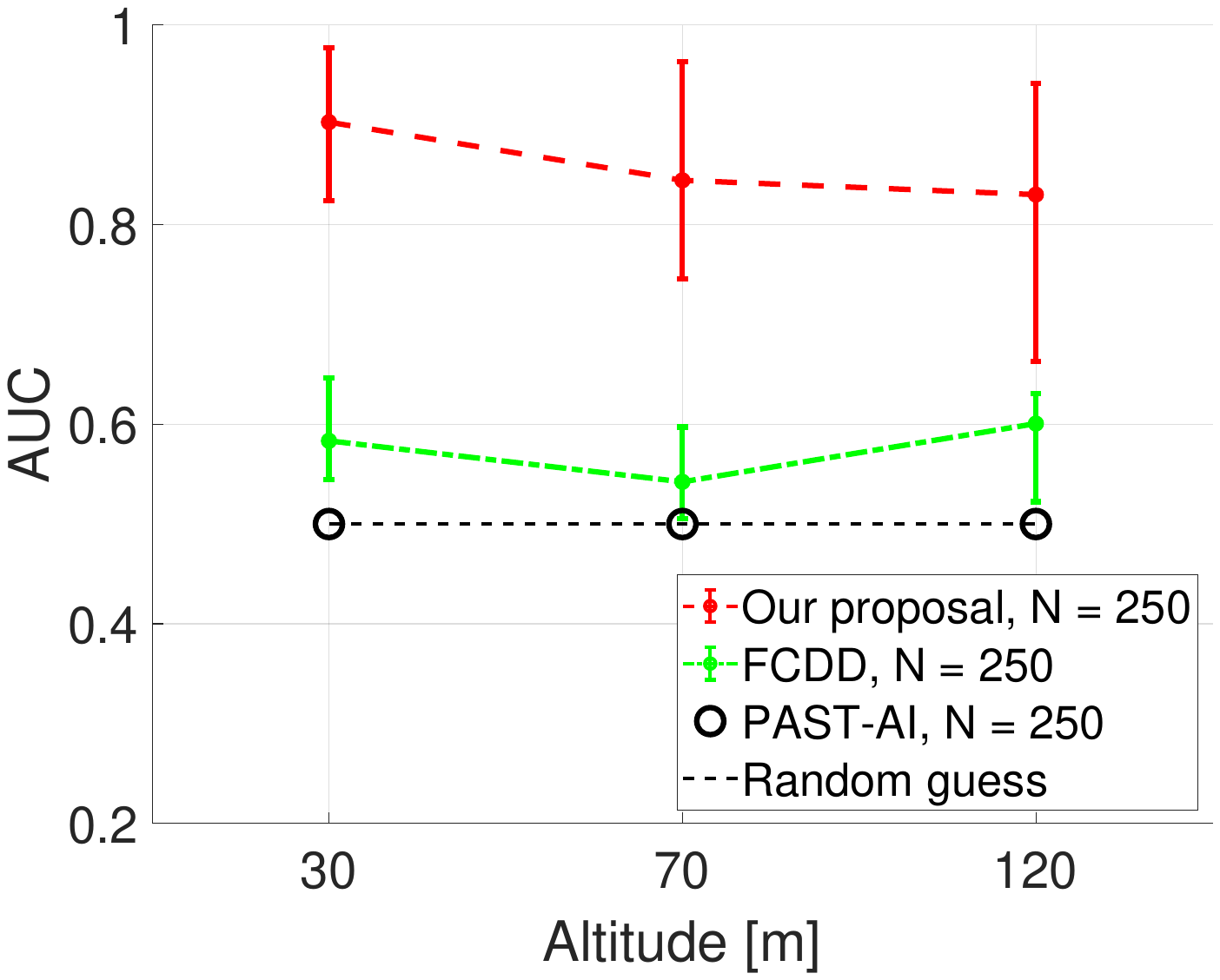}
        \caption{S3.}
        \label{fig:results_e3}
    \end{subfigure}
    \caption{
    Performance of our solution and the benchmarks in~\cite{oligeri2024sac} and~\cite{oligeri2023tifs} in scenarios S1 (a), S2 (b), and S3 (C), using $N = 250$ samples per image.
    }
    \label{fig:comparison}
\end{figure*}
We observe that our proposed approach consistently outperforms the benchmark solutions in all conditions. In fact, in all the investigated scenarios and altitudes, the two benchmarks rarely exceed a ROC AUC value of $0.6$, indicating that they can hardly detect spoofing activity. The solution using \emph{Resnet-18} proposed in~\cite{oligeri2023tifs} is the least effective, being characterized by an average accuracy equivalent to random guess, i.e., $ROC AUC=0.5$. This is due to the reduced number of samples used to build the corresponding images. Indeed, as highlighted by the authors in~\cite{oligeri2023tifs}, PAST-AI requires approximately $10^6$ samples to perform reliably, which is not feasible when performing spoofing attack detection.} The solution using the FCDD network proposed in~\cite{oligeri2020_wisec} is usually slightly more effective than the previous one but less effective than our proposed approach. We notice only one experiment where the solution in~\cite{oligeri2024sac} achieves slightly higher performance, i.e., S1 with the drone spoofing at $90$~m. Comparing Fig.~\ref{fig:comparison}(a) with Fig.~\ref{fig:results_autoencoder}(a), we can see that the performance of the benchmark aligns with the one achieved by our solution using $N=1,000$ samples per image; we believe that this result is due to the binary nature of the FCDD used in~\cite{oligeri2024sac}, which could potentially better discriminate spoofed from legitimate samples, since it is exposed to both classes during training. However, as mentioned earlier, collecting attack samples for the training process is often infeasible due to the required technical skills, expensive equipment, and law regulations. Our solution, using fewer samples per image, can provide similar results while requiring only regular satellite data (no anomalous samples at testing time). This feature further emphasizes the practical value of the proposed approach. Moreover, note that (aerial) spoofing can occur via various scenarios and channel conditions, being quite hard to model precisely through a binary classifier. Our proposed approach, requiring modeling only the regular profile of the received messages, is often even more successful than binary classification approaches in detecting anomalies.

\textcolor{black}{
\section{Discussion}
\label{sec:discussion}
\noindent
{\bf Implications for Practice.} The results provided in Sec.~\ref{sec:results} indicate that, through our solution, we can detect spoofing attacks to LEO satellite communication systems originating by drones by analyzing the physical-layer signals at the satellite receiver, and looking for anomalies in the received signal. At the same time, especially when considering $N=50$ samples per image, the average optimal ROC AUC values sometimes do not exceed the value 0.75. While such performances are considerably higher than the state of the art (see Fig.~\ref{fig:comparison}), they are not high enough to provide reliable detection in real-world systems. Thus, we foresee that our solution could be used to provide an additional layer of security to existing security systems, rather than as a standalone solution. This is common to many security systems analyzing wireless signals, e.g., Radio Frequency Fingerprinting~\cite{oligeri2023tifs}. We also notice that detection accuracy improves when there are more samples available per message and when the receiver is mobile (see Fig.~\ref{fig:comparison} (c)). Thus, whenever possible, receivers should oversample the incoming signal and move periodically to maximize detection accuracy and reduce false positives. 
\\
{\bf Limitations.} The results reported in Fig.~\ref{fig:results_autoencoder} and Fig.~\ref{fig:comparison} suggest that the attacker can increase the chances of bypassing our solution by flying higher, so better mimicking the satellite communication channel. However, due to the regulations in place in our region, in this research, we conducted tests with a drone flying at the maximum altitude of $120$~m---being the maximum allowed altitude~\cite{ilent_nl}. We acknowledge that attackers could neglect such regulations and execute advanced spoofing strategies by flying the drone at a higher altitude. Future research could test our solution with data collected in regions with more relaxed regulations, investigating further the capabilities of advanced attackers. Our tests were also executed in large rural areas, so to minimize safety concerns. Future research could test our solution with data collected in urban environments, to check if reflections caused by man-made constructions could help the attacker, or strengthen the defense strategy. 
}

\section{Conclusion and Future Work}
\label{sec:conclusion}

In this paper, we proposed a new solution for detecting aerial spoofing attacks to LEO satellite communication systems, using a sparse \acl{AE}. Our proposed solution requires converting received \acl{PHY} information (\ac{IQ} samples) into images and then using an \acl{AE} to identify which of them come from the spoofer. We validated our solution using data collected through an extensive real-world measurement campaign, encompassing actual spoofing attacks of a reference LEO satellite system (IRIDIUM) via a \acl{SDR} installed on a drone and flown in a remote area via several attack scenarios (static, transmitter movement, transmitter and receiver movement). Our proposed solution can detect aerial spoofing attacks with a minimum ROC \ac{AUC} value of $0.75$, systematically outperforming competing approaches. We also released the dataset collected for this study as open source to the research community, contributing to stimulating further real-world applied research on satellite security. 
\textcolor{black}{Note that, in our work, we could not fly the drone beyond the maximum altitude of $120$~m imposed by EU regulations. Also, we did not carry out tests in an urban scenario, where the presence of man-made construction and people' movement might change the profile of the signal received from satellites. In line with such limitations, in future work, we plan to test our solution in urban environments, so to assess the impact of man-made constructions on the effectiveness of our approach. Moreover, we will work on an analytical channel model for LEO satellite transmissions to analyze the performance of our solution in the presence of attackers circumventing law regulations on the maximum allowed flight altitude of drones. Finally, we will investigate the feasibility of detecting advanced drone-based LEO spoofing attacks onboard other drones and LEO satellites.}

\bibliographystyle{IEEEtran}
\balance
\bibliography{refs}

\end{document}